\documentclass[10pt]{article}
\usepackage{amsmath, a4wide, amssymb, mathrsfs, bbm, amsthm, chicago,url}
\usepackage[english]{babel}
\usepackage[mathcal]{eucal}
\usepackage[title]{appendix}

\newcommand{\real}[1]{\ensuremath{\mathbb{R}^{#1}}}

\newcommand{\integer}[1]{\ensuremath{\mathbb{Z}_{#1}}}

\newcommand{\vect}[2]{\ensuremath{{\mathbf #1}_{#2}}}

\title{{\Large \bfseries Electroweak Symmetry Breaking, Intermediate Regulators and Physics Beyond the Standard Model}}
\author{\normalsize \itshape Marc Holman\thanks{\itshape E-mail : m.holman@uu.nl},\\ \normalsize \itshape Department of Mathematics, Utrecht University,\\ \normalsize \itshape P.O. Box 80.010,  3508 TA  Utrecht, The Netherlands}
\date{}

\begin{document}
\maketitle 
\begin{abstract}
\noindent According to the long-standing received wisdom,  a ``small'' value of the Higgs mass - as for instance implied by 
general unitarity constraints - is highly ``unnatural'' and essentially \emph{requires} new physics to be present at or near
currently accessible energy scales.
Following the discovery of a new, Higgslike boson at the LHC facility in 2012, but with no sign of new physics
after having explored a large region of parameter space, a dilemma thus seems to present itself :
either the newly discovered boson is indeed the long-sought Higgs boson of the standard model of particle physics
(or some appropriate variant of that model) and the new physics at the TeV scale, supposedly required by the naturalness 
argument, is still waiting to be discovered, possibly by LHC-II, 
or the identification of the new boson as the Higgs cannot be maintained.
It is shown that this apparent dilemma is in fact a false one, in that nothing in contemporary particle physics 
dictates that a small Higgs mass be unnatural in any way.
\end{abstract}


\section{Introduction}\label{intro}

\noindent It now seems almost unimaginable that modern science has witnessed a brief episode in which
the entire material world could in principle be understood on the basis of just three elementary particles.
Yet, it was only in 1928, just a few years after quantum theory had been born, when physics held it that all known 
forms of matter were in principle reducible to complex collections
of interacting electrons, protons and photons and when the time was apparently not ripe to postulate the existence 
of even a \emph{single} new particle - in spite of the fact that such a particle, i.e., the positron, was a straightforward
prediction of a brand-new relativistic theory of the electron, highly successful in every other respect.
Of course, there were also numerous, well-known problems with this simple picture, which enforced substantial adjustments 
of it shortly afterwards : new forms of matter were soon postulated on theoretical grounds (i.e., the electron
neutrino in 1930, the positron, after all, in 1931, dark matter in 1933, pions in 1935), as well as experimentally
discovered (i.e., the neutron in 1932, the positron, not initially identified as such, in the same year) and in the wake
of these adjustments the number of subatomic particles has grown and grown - probably beyond even the wildest imaginations 
of any physicist back in 1928.
Although great progress was made in the 1970s, when it was shown that all the particle trophies that had been
brought back from experiments since the 1930s could be explained in terms of a relatively straightforward model that 
postulated a ``mere'' twenty-five elementary constituents, it was also clear from the outset that this so-called 
standard model left too many questions unanswered.\\
Now that it appears that all basic constituents of the standard model have finally been observed in experiments
- some four decades after the model's birth - it may therefore seem appropriate to try and take stock.
One is inclined to ask, what are the lessons of the standard model ?
Does the model provide any clues about how to extend physics beyond it ?
At a deeper level, what does it teach us, after Gauguin's painting, about where we come from, what we are and where we are going ?
Although physicists unfortunately are sharply divided over the probable answers to these questions, there does
appear to be broad agreement that some radical renewal, both in concepts and theory, is required in order to move physics 
beyond the standard model.
In this regard, the newly discovered boson with mass
\begin{equation}\label{Xbosonmass}
M_X \; = \; 125 \, \mbox{GeV}
\end{equation}
that was announced by the CERN laboratory on July 4th 2012, will likely mark a watershed for particle physics\footnote{\citeN{ATLAS}, \citeN{CMS}.}.
If the new boson indeed turns out to be the long-sought standard model Higgs boson - as most particle physicists now seem to believe\footnote{Although the
detection of the new boson has been widely advertized in the media as the long-awaited discovery of the Higgs boson and has
even already been honoured as such in the form of various awards (most notably the 2013 Physics Nobel Prize), it has
been pointed out that these developments were actually premature. See \citeN{Moffat}.
In fact, the underlying problem already becomes clear from the official CERN viewpoint itself.
In a public report of October 2011, the \citeANP{CERN} explicitly states that in case of a discovery of a state
consistent with the standard model Higgs boson, it would still be necessary to check \emph{with the greatest achievable
precision whether or not the observed state has indeed the properties expected of a Higgs boson in the standard model}
(i.e., through verification of its charge, mass, spin, parity, couplings to the weak vector bosons and fermions, etc.) and that this
\emph{would require many years of operation of the LHC, accumulating the largest possible data set at the highest achievable
energy} (italics added).
Since the LHC was shut down for maintenance in February 2013 and did not resume operation until June 2015, this makes the 
timing of the mentioned awards quite remarkable.} - this should in principle and for the first time since its conception
place theorists in a very secure position to assess the invariant physical content of the 
standard model - considered as a closed theoretical construct, unambiguously corroborated by experiments.\\
Quite remarkably, the Higgs boson and associated mechanism have always had a somewhat odd status within the
standard model. As is well known, the Higgs mechanism was proposed by a number of authors independently in the mid 1960s, as a means
to ``break'' gauge symmetry and thereby ``supply'' masses to gauge bosons\footnote{Cf. \citeANP{Higgs1} \citeyear{Higgs1,Higgs2}, \citeN{EngBro}, \citeN{GuHaKi}.\label{Higgsrefnote}}.
Concretely implemented within the Glashow-Salam-Weinberg model, i.e., the ``electroweak sector'' of the standard model,
the mechanism effectuates a transition from a phase with a manifest $\mbox{SU}(2)\times\mbox{U}(1)_{\mbox{\scriptsize Y}}$-symmetry
(generated by weak isospin and weak hypercharge, respectively), to a phase with a manifest $\mbox{U}(1)$-symmetry
(generated by ordinary electric charge), effectively leaving three massive weak vector bosons, $W^{\pm}$, $Z^0$
and one massless photon, $\gamma$.
However, although the standard model (at least in its modern formulation) also predicts the existence of a massive
spin-$0$ particle as a consequence of this phase transition, i.e., the Higgs boson, it has virtually nothing to say
about the properties of this putative particle\footnote{Although the Higgs mechanism fixes the masses of 
both the vector bosons and the Higgs boson in terms of other parameters (at tree level), in the latter case, one of these
parameters (the scalar self-coupling) is unknown and is only constrained by general stability requirements, which are
not very informative. For instance, in the mid 1970s, the condition that the vacuum be stable only implied that the
Higgs mass had to be larger than $3.72 \, \mbox{GeV}$. Cf. \citeN{Weinberg8} (later it was realized that this lower
bound significantly increases in the presence of elementary fermions with a mass exceeding $\gtrsim 80 \, \mbox{GeV}$).
The fact that the standard model provides little information about the properties of its Higgs boson is underscored
by the fact that alternative, empirically viable models of electroweak symmetry breaking exist in which Higgs bosons have no place
(see e.g. \citeN{Ferrari}, \citeN{Moffat}).}.
Furthermore, it has long been recognized that the mechanism itself is conceptually problematic because of its gauge dependence\footnote{Cf. \citeANP{Earman1} \citeyear{Earman1,Earman2}.},
while it was shown more recently that in its standard textbook formulation, the Higgs mechanism is in fact conceptually inconsistent\footnote{Cf. \citeN{Holman1}.
For a different argument leading to the same conclusion, see \citeN{GuHa}.}.
Although it appears that these difficulties can be overcome, doing so would probably also compromise specific features
of the standard formulation which are often regarded as crucial (such as locality or renormalizability).\\
A rather different peculiarity of the standard model Higgs boson - one that has in particular been stressed by
particle physicists for many years - is a bit more technical and finds its origins in the radiative corrections
to the Higgs mass in perturbation theory.
In contrast to analogous corrections to fermion and gauge boson masses, which only display a ``logarithmic'' divergent 
behaviour, radiative corrections to the Higgs mass depend quadratically on an explicit high-energy cutoff, $\Lambda$, 
and this is typically argued to render the ``natural'' value of the Higgs mass of order $\Lambda$.
Since, assuming that the Higgs indeed exists, a ``best fit'' based on electroweak precision data yields a value for its mass
near $90 \, \mbox{GeV}$ (with the value (\ref{Xbosonmass}) falling just outside the uncertainty range)\footnote{\citeN{alephcs}.}, 
while general arguments are taken to imply that the Higgs mass can at most be an order of magnitude larger,
these estimates would thus be inconsistent with the natural mass value supposedly implied by general quantum field 
theoretical considerations, if $\Lambda$ is beyond TeV order.
In arguments of this type, $\Lambda$ is taken to represent some sort of ``physical'' cutoff to the standard model
and the inference to be drawn is that if the newly found boson is indeed the Higgs (or, more generally, if a Higgs
exists in the expected mass range), naturalness \emph{requires} new physics to be lurking near the TeV scale.
One possibility sometimes entertained in this regard, for instance, is that the Higgs boson is a composite particle bound together
by very strong ``technicolour'' interactions at low energies, somewhat akin to a Cooper-pair of electrons in the superconducting
state.
However, for many years, the most popular new physics proposal by far in this regard has been based on the postulated existence
of a novel kind of symmetry, a so-called ``supersymmetry'', which essentially relates every standard model boson and 
fermion to a hithertho unobserved ``superpartner'' of opposite statistics, i.e., a fermionic ``bosino'' and bosonic 
``sfermion'', respectively.
The presence of a Higgsino is then supposed to ``stabilize'' the Higgs mass, by effectively lowering its superficial
degree of divergence.\\
Although the prospect of new physics hiding just behind the corner, so to say, is of course exciting (but cf. section \ref{disc}), 
it is somewhat surprising that the argument that is supposed to unquestionably lead to a conclusion of this extent has been
scrutinized very little so far. It is the purpose of this article to fill this gap.
To this end, a conceptual, but technically accurate, short review of perturbative renormalization in quantum field theory
and the problems it leaves unaddressed, is first presented in section \ref{QFTinfinities}.
This is then followed by a critical discussion of the modern perspective on these problems in terms of effective field 
theories in section \ref{SMtriv}.
Although the discussion in these two sections is general and may at first glance seem unrelated to the naturalness
arguments based on estimates for the Higgs mass, as will be seen in section \ref{naturalness}, the whole issue of 
renormalization is in fact of crucial significance here.
As a matter of fact, it will be seen that a Higgs naturalness problem for the standard model simply does not exist
and that the apparently received view to the contrary finds its origin in an incoherent application of renormalization theory.

\section{Field Theoretical Infinities and Their Intended Remedies}\label{QFTinfinities}

\noindent The impressive empirical adequacy of the standard model is at the same time a bit embarrassing, because of the 
long-standing difficulties to make mathematical sense of the quantum field theoretical basis of that model, i.e.,
as a consequence, essentially, of the unavoidable appearance of certain divergences in it.
According to the old orthodoxy, these divergences can be successfully removed through a procedure called \emph{renormalization},
which is precisely effective because standard model quantum field theories happen to belong to a special class of theories
- namely those characterized by the absence of dimensionful coupling constants.
According to contemporary orthodoxy however, the mathematical consistency of renormalized quantum field theories not only
remains to be proven, but is very probably even elusive, strictly speaking - hence, the slightly embarrassing agreement 
between theory and experiment in high-energy physics.
Nevertheless, in spite of its somewhat disreputable status, it is worth stressing that there is nothing in itself
illegitimate about the act of ``absorbing infinities'' into arbitrary parameters present within the formalism,
which, in one way or another, is an essential part of any renormalization procedure (and not infrequently is
even identified with it).
A good way to illustrate this - also as a warm-up excercise to the technically more challenging situation in
quantum field theory - is in terms of the infinite ``self-energy'' of a classical point electron.

\subsection{Mass Renormalization in Classical Electrodynamics}\label{CEDrenorm}

\noindent As is well known, considerations about the nature of electrically charged particles - in particular questions 
regarding their extension and self-field effects - played an important role in the development of various extended electron models 
at the dawn of the twentieth century\footnote{Cf. \citeN{Jackson}, \citeN{Rohrlich}.}.
A key feature of these approaches was that they attempted to describe the electron (or in fact any charged particle) in purely 
electromagnetic terms. In particular, the electron's mass was to be understood as a purely electromagnetic effect. 
To see the rationale for this, recall that the electrostatic potential energy of a uniform, spherically symmetric charge distribution of 
total charge $q$ and radius $\xi$ is given by 
\begin{equation}\label{statpotential}
U_{\xi} \; = \;  \frac{q^2}{8 \pi \xi} \biggl( 1 \: + \: \frac{1}{\alpha} \biggr)
\end{equation}
where in the case of a spherical shell, respectively solid sphere, $\alpha = \infty$, respectively, $\alpha = 5$. 
From an application of Newton's second law and the usual Lorentz force law, it can then be inferred that, for an 
electron moving uniformly with non-relativistic velocity, \vect{v}{}, and up to an arbitrary constant, the electron's 
momentum, $\vect{P}{}$, equals minus the ``field momentum'', $\vect{P}{\mbox{\footnotesize field}}$, i.e. minus the 
spatial integral over the electromagnetic momentum density, $\vect{g}{} = \vect{E}{} \times \vect{B}{}$, which for 
an electron that has its charge, $e$, uniformly distributed over a spherical shell of radius $\xi$, is given by
\begin{equation}\label{EMfieldmom}
\vect{P}{\mbox{\footnotesize field}} = - \frac{e^2}{6 \pi \xi} \vect{v}{} = - \frac{4}{3} U_{\xi} \vect{v}{}
\end{equation}
with $U_{\xi}$ given by (\ref{statpotential}) (for $\alpha^{-1}=0$ and $q=e$).
Upon setting the arbitrary constant equal to zero, this then suggests to identify $4 U_{\xi} / 3$ with the (rest) mass, $m_e$, of the electron\footnote{The issue of whether actually $U_{\xi}$ or $4/3$ times this quantity should be identified with $m_e$ is a bit
of a quagmire. Historically, the first option was chosen by Abraham and Lorentz in their model of the electron, but this led
to an anomalous coefficient of $4/3$ for $m_e$ in the equation of motion, which was subsequently attributed to the relativistic
noncovariance of the model. It was then argued by Poincar\'e that, in order to maintain relativistic covariance, as well as
stability, it is necessary to introduce additional stresses, partly of non-electromagnetic origins, for the extended 
electron model. However, within a relativistically covariant context, Fermi subsequently 
found that the appearance of the factor $4/3$ is due to an incorrect definition of the field momentum as a spatial 
integral over the momentum density \vect{g}{}, implying that it is actually unrelated to the Poincar\'e stresses. For a 
detailed account of these issues, see \citeN{Rohrlich}.\label{xtndelectron}}.
However, a major difficulty facing the extended electron model as just characterized, is what keeps the electron from 
falling apart (i.e., as emphasized by Poincar\'e, non-electromagnetic interactions are required for stability; cf. footnote \ref{xtndelectron}).
On the other hand, in the limit $\xi \rightarrow 0$, no stabilizing interactions are required, but $U_{\xi}$ - and hence
the electron mass, if viewed in entirely electromagnetic terms - clearly diverges.
Either way, these remarks imply that a purely electromagnetic account of mass is not a viable option.\\
In typical discussions, the fact that $U_{\xi}$ is linearly divergent as $\xi \rightarrow 0$ is characterized by saying that
``the self-energy of a classical point charge is infinite''.
It is because $U_{\xi}$ is also nontrivially related to a physical observable, viz., mass, that the renormalization 
procedure enters classical electrodynamics\footnote{It should be noted however that the case for renormalization in 
classical electrodynamics is a bit more subtle than just presented.
First, although the infinite nature of the ``self-energy'' of a classical point charge seems reasonable if that charge
is viewed as the result of collecting together - from infinity or otherwise - a distribution of ``infinitesimal charges'', 
it seems impossible to think of such a process as being in principle realizable in any finite amount
of time (i.e., even in a strictly classical world, it would require increasingly more work with each ``collapse'' of the
distribution by any prescribed, finite amount, to overcome electrostatic repulsion).
Yet, it should clearly require no work to bring a point charge - viewed as a single, indivisible entity - to 
any specified position in an otherwise empty spacetime, entailing that the ``self-energy'' of such a charge actually 
vanishes (one is of course dealing here with a highly idealized thought experiment, but if initial and final 
non-uniformities in the motion are neglected, no work is done by uniformly moving the charge through its radial 
Coulomb field).
For such reasons, it seems more reasonable to attribute the divergence in the ``self-energy'' of a point charge to
an an illegitimate use of concepts. See also \citeN{Frenkel}.
On the other hand, as also seen in the text, $U_{\xi}$ is nontrivially related to the physical mass of a charge
distribution and even though the argument that establishes this relation is valid only for finite $\xi$, it is
desirable to be able to treat the case of a point charge as being continuously retrievable from the case of a finite
charge distribution.
This is especially so when it is attempted to derive a dynamical equation for point charges in which ``radiative reaction'' 
effects are taken into account (not surprisingly, the potential energy of an initially finite charge distribution
also enters here as a coefficient of the acceleration term; cf. \citeN{Dirac1} for a treatment of the relativistic case).
For further discussion of this point, see also \citeN{Holman2}.}.
Indeed, upon simply re-introducing the arbitrary constant set to zero in the previous discussion, it is easy to see that 
the observable, ``physical'' electron mass, $m_e \simeq 0.5 \, \mbox{MeV}$, can be expressed according to
\begin{equation}\label{CEDmassrenorm}
m_e \; = \; m_0 \: + \: \delta_{\xi} m
\end{equation}
where the ``mechanical mass'', $ m_0$, denotes an arbitrary integration constant and where $\delta_{\xi} m := 4 U_{\xi} / 3$.
But, clearly, $m_0$ is an arbitrary constant for \emph{each} possible value of the
electron radius $\xi$. Thus, by simply \emph{choosing} this constant so as to satisfy (\ref{CEDmassrenorm}) for each of these values,
the divergence in the observable, physical mass of the distribution can be ``absorbed'' as $\xi \rightarrow 0$
and one is left with a unified dynamical treatment of charges that is valid also in the point-like limit.
Although it is quite evident that this procedure of ``renormalization'' is completely ad hoc, there is nothing logically inconsistent 
about it (at this level of the discussion) and the foregoing example actually provides a good illustration of what a 
significant part of this procedure in practice amounts to in quantum field theory\footnote{It may of course be asked whether the account is \emph{physically} appropriate, but that 
is quite a different matter.}.
In fact, even though complications arise for the renormalization procedure in quantum field theory (both on a conceptual 
and a technical level) and even though, as noted, the mathematical consistency of this procedure so far remains to
be demonstrated, in the case of the ``self-energy'' of a point charge, there is at least one rather well-defined 
sense in which quantum field theory is better behaved than classical field theory, since, according to quantum electrodynamics 
(QED), the self-energy of a point charge diverges only logarithmically (as a function of momentum, or, essentially equivalently, inverse length).

\subsection{Operator Fields, Locality and Renormalization Theory}\label{renormalization}

\noindent In quantum theories of fields, divergences also appear in expressions for particle self-energies and according
to renormalization theory, it is possible to remove these divergences through redefinitions of physical observables,
such as mass and charge, essentially just as in classical electrodynamics (i.e., in the case of QED, the photon mass
necessarily vanishes as a consequence of the $\mbox{U}(1)$ gauge symmetry - according to the standard view on these
matters at least - and the photon self-energy, i.e., ``vacuum polarization'', is absorbed into the definition of the
physical electron \emph{charge}).
Furthermore, just as in classical electrodynamics, self-energy divergences in quantum field theories essentially are a
direct consequence of the point-like status of ``particles'' in such theories.
There are, however, a number of subtleties and marked differences as compared to the classical situation.
First, as indicated by its name, a quantum theory of fields is indeed fundamentally a theory of \emph{fields}, not
particles\footnote{This issue is neither academic nor semantic.
In quantum field theory in curved spacetime, there is in fact no preferred notion of ``positive frequency solutions''
(or, equivalently, a preferred vacuum state) and consequently, there is no natural ``particle'' notion (this is a 
direct consequence of the fact that infinitely many unitarily inequivalent representations of the canonical
commutation relations in general exist and that, in a general curved spacetime, there is no way for singling out
a preferred representation; see e.g. \citeN{Wald}).
But even for quantum field theory in Minkowski spacetime (where the Poincar\'e symmetry makes it possible to pick out
a preferred space of positive frequency solutions), the fact that the standard vacuum state will appear as a thermal
state to a uniformly accelerating observer (\citeN{Unruh}), or the fact that in those circumstances where they can be
defined, ``particles'', i.e., field-``quanta'', are highly nonlocal entities which moreover defy a local realist interpretation on the basis of general no-go results,
makes it clear that the notion of ``particle'' does not have objective significance (independent of the question of 
whether quantum field theories are valid as ultimate descriptions of physical reality).
Hence, even for quantum field theory in Minkowski spacetime (which will be the general context assumed further throughout 
this article), any talk of (point-like) ``particles'' is useful for metaphoric purposes at most.\label{QFTsubtleties}}.
Second, in many quantum field theories of physical interest, ``elementary'' divergences also arise in expressions
not directly relatable to the masses or coupling constants of field quanta (this occurs, for instance, for the basic
electron-photon vertex in QED). In fact, the situation is somewhat more complicated still, since divergences can also
overlap, as a result of which redefinitions of field variables are typically also required (this occurs, for 
instance, for the self-energy in $\lambda \phi^4$-theory; cf. subsection \ref{phi4renorm}).
What can be said however is that for present purposes, the quantum field theoretical infinities (in the ultraviolet)
addressed by renormalization theory all appear in the form of expressions involving products of field operators at
the same spacetime point, which is itself a direct a consequence of the locality assumption underlying general quantum 
field theory\footnote{There are again some subtleties. First, already in the case of a free field, even though a solution
to the field equation in terms of an expansion into normal modes can formally be written down (upon ``putting the field in a box'' subject
to periodic boundary conditions), the resulting series does not converge and it is in general necessary to ``smear''
quantum fields with test functions - thereby turning them into ``operator valued distributions'' - in order that they
be mathematically well defined (it should incidentally be stressed that this is only a mathematical step, i.e., to
give rigorous meaning to operator fields, which does not essentially change the conceptual model for quantum field theory 
in terms of pointlike operator fields and local interactions; see also below).
Second, as discussed in the main text, the divergences addressed by renormalization theory all appear in the context
of perturbation theory and there are strong indications that renormalized perturbation series for physically relevant
theories such as QED are probably not convergent, while there moreover appear to be formidable obstructions (so-called
``renormalons''), which are related to the general consistency question of renormalization proofs, to consistently 
``resumming'' such series. See e.g. \citeN{Wightman} for a clear discussion of this point.
In fact, for asymptotically free theories such as QCD, divergence is also expected, but opinions seem to diverge as 
to whether the singularities appearing in Borel resummed series are harmful or not (cf. \citeN{tHooft4}).
Finally, regarding the locality assumption, there are two (not unrelated) aspects to this assumption : local (i.e., 
point-like) operator fields and local interactions (i.e., rejection of action at a distance).
Given the first aspect, it is the second aspect that entails the need to take field operator products evaluated at the same spacetime point.
See e.g. \citeN{Cao}, section 3, for further discussion of these matters.\label{fieldsubtleties}}.
For free fields, which can be rigorously defined as operator-valued distributions, there exists a canonical
subtraction procedure for unambiguously removing these infinities (which appear in nonlinear observables, such as energy), called 
\emph{normal} (or \emph{Wick}) \emph{ordering}.
For interacting fields, however, no canonical subtraction procedure is known to exist, but there does exist a general, 
in principle reasonably clear-cut, heuristic scheme for perturbatively ``removing'' field theoretical infinities, which, 
in some important cases, has compared very favourably to experiment.\\
Now, without going into the full technical machinery of this scheme - which is often generically referred to as
``renormalization theory'' - the basic idea here is to assume that heuristic meaning can be given to ``interacting
Green's functions'' (i.e., ``full vacuum'' expectation values of time-ordered products of field operators in the 
Heisenberg picture) and that, for sufficiently weak interactions, these functions can be perturbatively expressed
in terms of ``free'' Green's functions (i.e., essentially ``free vacuum'' expectation values of time-ordered products
of field operators obeying free field equations).
Since, the latter can always be expressed as sums of products of free two-point Green's functions (i.e., Feynman amplitudes) 
by virtue of Wick's theorem, it would appear from the previous remarks that the only nontrivial issue that remains is the 
convergence of the perturbative series as a whole.
That this is in fact not the case becomes clear upon inspecting the explicit formula for a general, interacting
$n$-point Green's function, obtained through the above heuristic reasoning\footnote{Cf. \citeN{GelLow0}.}
\begin{equation}\label{GLformula}
G (x_1 , \cdots , x_n) \; =  \; \langle \mbox{T}\{ \Phi(x_1) \cdots \Phi(x_n) \exp \left( i \int \! d^4x \, {\mathscr{L}}_{\mbox{\scriptsize int}}(x)\right) \} \rangle_{\mbox{\scriptsize conn}}
\end{equation}
where $\Phi$ denotes a generic quantum field in the ``interaction picture'' (i.e., obeying, somewhat counter-intuitively, 
a free field equation), $\mathscr{L}_{\mbox{\scriptsize int}}$ denotes the part of the Lagrangian density involving 
interactions, $\mbox{T}$ denotes the usual time-ordering symbol and ``conn'' designates that eq. (\ref{GLformula}) 
contains no ``vacuum bubbles'' unconnected to any of the $n$ ``external'' points.
The problem is not, as is sometimes thought, merely the infinitely many terms implicit in eq. (\ref{GLformula}) that involve
divergent products of field operators evaluated at the same spacetime point (i.e., as noted above, for free fields,
such divergent products can be dealt with non-ambiguously through normal ordering).
The problem is rather that, upon assuming that the normal ordered forms of these products (which are usually referred
to as \emph{Wick polynomials}) have been turned into rigorously defined distributional operator fields, the discontinuity 
of the step functions (implicit in the time ordering) in (\ref{GLformula}) introduces ambiguities.
That is, the general notion of a product of a distribution with a discontinuous function can easily be shown to
lead to contradictions and thus lacks mathematical content.\\
Nevertheless,  the right-hand side of eq. (\ref{GLformula}) can be formally expanded as a power series in the coupling constant(s) implicit in
$\mathscr{L}_{\mbox{\scriptsize int}}$ (in fact, the exponential function in (\ref{GLformula}) is only \emph{defined}
as a formal series) and this leads directly to the familiar perturbative analysis of a quantum field theory in terms
of Feynman diagrams.
Now, again without going into the technical details, there are four crucial steps involved in standard arguments
for extracting empirical content from a quantum theory of fields, based on the generic, formal power series (\ref{GLformula}) :
\begin{itemize}
\item[(i)] Simple dimensional analysis leads to a direct classification of interactions according to potential physical
viability. That is, interactions with coupling constants of negative mass dimensions basically display a more violent
short-distance behaviour with each increasing order in perturbation theory and this means (again, according to a
standard line of reasoning), that no physically meaningful information can be extracted from them on the basis of 
(\ref{GLformula}). Such interactions (or the theories in which they appear) are therefore called (perturbatively)
\emph{non-renormalizable}.
On the other hand, the short-distance behaviour of interactions with couplings of non-negative mass dimensions is
radically different, in that it does not become more violent (and actually even improves for strictly dimensionful 
couplings) with each increasing perturbative order and this is taken to mean that physically meaningful information 
can in principle be extracted from (\ref{GLformula}) for such interactions.
Interactions of this type (or those theories in which they appear exclusively) are therefore called (perturbatively) \emph{renormalizable}
(and, in fact, even ``super-renormalizable'' in the strictly dimensionful case)\footnote{The significance of the dimensionality
of coupling constants can be understood as follows. 
A rough estimate on the short-distance behaviour of any term (i.e., any Feynman diagram) in a formal series (\ref{GLformula})
generalized to $d$ dimensions, follows from naive ``power counting''. That is, a simple counting of powers
of momenta associated with different pieces of a given (Fourier-transformed) term, $\Gamma$, directly leads to its
so-called \emph{superficial degree of divergence}, $D_{\Gamma}$, which can be expressed in the following general form
\begin{equation}\label{superficial}
D_{\Gamma} \; = \; d \: - \: \sum_{\alpha}(d - \delta_{\alpha})V_{\alpha} \: - \: \sum_i d_i E_i
\end{equation}
Here, the indices $\alpha$, $i$ respectively label the different possible vertices and fields of the theory,
$V_{\alpha}$ denotes the number of vertices of type $\alpha$, $E_i$ denotes the number of ``external lines''
of type $i$, $d_i$ denotes the mass-dimension of a type-$i$ field (which, except for non-standard theories
involving higher-order derivative kinetic terms in the action, is positive) and $d - \delta_{\alpha}$, finally,
denotes the dimension of the coupling constant associated with a type-$\alpha$ vertex (since issues such as 
renormalizability, the existence of renormalization group fixed points, and so on, turn out to have a crucial
dependence on the dimension of the spacetime model on which a given quantum field theory is defined, it is often 
convenient in discussions to consider a slightly more generalized setting, with the dimension of spacetime, $d$, 
treated as variable).
Now, it is immediately clear from eq. (\ref{superficial}) that for theories involving only couplings of non-negative
dimensions, superficially divergent diagrams arise only for a finite number of $n$-point functions : for $n$ sufficiently
large, all Feynman diagrams appearing in the series expansion (\ref{GLformula}) of such a theory  will be superficially convergent (i.e., have $D_{\Gamma} < 0$).
However, contrary to what is suggested by (\ref{superficial}), even these latter diagrams will still be divergent
in actuality if they contain divergent subdiagrams. Hence, the adjective ``superficial''.
Nevertheless, according to renormalization theory, the fact that only finitely many $n$-point functions contain 
superficially divergent diagrams is of crucial importance, in that the removal of divergences from these functions
will automatically remove divergences also from all other $n$-point functions.
By contrast, in the case of interactions with coupling constants of negative dimensions, eq. (\ref{superficial}) shows 
that at high-enough order in perturbation theory, every Feynman diagram corresponding to a given physical process is 
superficially divergent.
In other words, \emph{every} $n$-point function contains infinitely many superficially divergent diagrams, while
the divergent behaviour moreover becomes worse with each subsequent order in perturbation theory.
It should be noted however that a classification of theories based on the notion of renormalizability, although 
still useful for practical purposes, has become somewhat obsolete and a more modern classification is based on
the notion of \emph{asymptotic safety} (see subsection \ref{QFTperspectives}).
According to this modern view, a non-renormalizable theory is in principle viable if its beta-function posesses 
a nontrivial zero. In fact, explicit solutions to such theories, involving only finitely many arbitrary parameters are known to
exist. See e.g. \citeN{Wightman}.\label{supdivnote}}.
\item[(ii)] For any renormalizable theory, even though all non-vanishing Green's functions contain ambiguities (i.e., for
each $n$, (\ref{GLformula}) contains ambiguities unless $G=0$ identically), detailed technical arguments are taken to
imply that all these ambiguities will have been removed if and only if all ambiguities will have been removed from a finite
number of Green's functions (cf. footnote \ref{supdivnote}) and that, moreover, this will have been achieved if and only if for each of
these latter Green's functions, all ambiguities will have been removed from a special (infinite) subclass of terms in
its representation (\ref{GLformula}). In terms of Feynman diagrams, this special subclass is just the set of all \emph{proper} diagrams corresponding to that 
particular Green's function\footnote{For 
any value of $n$, the series (\ref{GLformula}) contains many diagrams that are physically redundant, in the sense that they do not contribute to the nontrivial 
part of the $S$-matrix. Technically, the only physically relevant diagrams in (\ref{GLformula}) are those for which all $n$ 
external lines are connected to each other and free from divergent contributions (i.e., ``internal loops''). This is
guaranteed by the so-called LSZ reduction formulae.
Furthermore, amongst all these ``fully connected'' and ``truncated'' diagrams, the only ones relevant for renormalization
are the so-called \emph{proper} or \emph{one-particle irreducible} (1PI) diagrams, i.e., the fully connected, truncated 
diagrams that remain connected after cutting a single internal line.
Proper diagrams (or more precisely, their analytical counterparts) are thus in a sense the elementary, perturbative building
blocks of general Green's functions (cf. also eq. (\ref{phi4-2point2})).}.
\item[(iii)] It is assumed that the perturbative assumption is indeed valid; i.e., even though any
coupling constants are themselves implicitly defined in terms of interacting Green's functions (and are thus strictly 
speaking infinite), it is assumed that in order to extract empirical information to any desired degree of accuracy,
for each proper Green's function only finitely many terms in its formal series expansion (\ref{GLformula}) need to be considered.
\item[(iv)] Justification for the perturbative assumption is sought in terms of the renormalization step.
That is, all ambiguities - which in practice appear in the form of divergent integrals in individual terms in the
series (\ref{GLformula}) - are ``absorbed'' into the finitely many, initially arbitrary (and hence ``physically undefined'') parameters
and fields present in the (complete) Lagrangian density. To this end, these ``bare'' parameters and fields
are themselves expressed as perturbation series in the physical, i.e., renormalized, coupling constant(s) - so that
(\ref{GLformula}) arguably acquires a more truly perturbative character - and are supposed to ``contain infinities'' 
at each order.
The key idea is then that \emph{to each order in perturbation theory}, all divergent integrals can be consistently cancelled
against the infinities contained in the bare parameters and fields to that order.
The formal proof that this is indeed possible precisely for renormalizable theories is highly nontrivial and is 
taken to be imperative for the physical viability of such theories.
Note however, that except for their perturbative nature, these redefinitions or renormalizations of parameters and
fields in quantum field theory are completely analogous to the previously considered example of mass renormalization
in classical electrodynamics\footnote{The appearance of perturbation theory in the quantum case of course amounts
to no small difference.
In particular, the step of perturbatively expanding the bare parameters into the renormalized coupling constant(s),
appears to render the entire argument circular (i.e., one does not know whether this step is legitimate \emph{until} the
removal of ambiguities to all orders has been proven and the latter crucially depends on the perturbative step),
whereas if the perturbation series do not converge, the assumption that a small part is being ignored is obviously
unjustified.\label{perturbdiff}}.
In typical discussions, the renormalization step is made more explicit by decomposing it into three sub-steps\footnote{It
is important to note that, within the context of the original renormalization theory that is considered here, such a 
decomposition is not necessary. Indeed, as exemplified by the theories of renormalization developped by Schwinger, 
Bogoliubov and others, the introduction of intermediate regulators is not \emph{required} for renormalization.
See also \citeN{CaoSch}. However, apart from its widespread use, the treatment of renormalization in terms of
regulators will turn out to be crucial for the discussion in section \ref{naturalness}.\label{nonnecintmedreg}} :
\begin{itemize}
\item[(a)] All divergent integrals are first \emph{regularized}.
That is, they are turned into finite integrals through the introduction of an additional parameter, i.e., a ``regulator'',
such that the original, divergent expressions are recovered for a certain limiting value of this parameter.
This step of implementing a particular regulator (or ``regularization scheme'') thus corresponds to an explicit modification 
of the theory and there are a priori many different ways to do so.
One, rather crude, way is to simply introduce an explicit high-energy \emph{cutoff}, $\Lambda$, as an upper integration limit
into the (Fourier-transformed) expressions. This has the advantage of being physically intuitive (at first glance, at
least) and it essentially figures prominently within Wilson's conception of the renormalization group discussed in section \ref{QFTperspectives}.
It does not, however, respect Lorentz and gauge invariance and, in fact, especially when it comes to Yang-Mills theories, 
a superior scheme is that of \emph{dimensional regularization}, in which the dimension of spacetime is modified by an 
infinitesimal amount $4 \rightarrow 4 - \epsilon$ (although far from obvious, this indeed renders the divergent 
integrals formally finite).
\item[(b)] To each order in perturbation theory, the parts of the evaluated, finite integrals in which the divergences
re-appear upon removing the regulator are to be absorbed into parameter- and field-redefinitions according to a particular
\emph{subtraction} (or \emph{renormalization}) prescription.
Again, there are many different possible choices for such a prescription.
One frequently used scheme (especially within the context of dimensional regularization, although its use is not restricted
to that context), is the so-called \emph{minimal subtraction} (MS) scheme, in which the ``counterterms'' (see subsection
\ref{phi4renorm}) are to \emph{just} cancel the dormant divergences.
\item[(c)] Finally, the regulator simply is to be removed in order to retrieve the original theory.
\end{itemize}
The representation of renormalization as the three-step procedure just outlined immediately raises the question of whether
the final, renormalized expressions depend on the choices for the regularization and subtraction scheme.
A quick glance at the literature suggests that in practice specific choices are effectuated mainly on the basis
of convenience. At any rate, it seems intuitively reasonable to expect that the ``physics'' should be independent
of either scheme.
However, regularization- and subtraction-independence of ``physics'' is a very broad criterion and it will be
seen later (in respectively sections \ref{naturalness} and \ref{SMtriv}) that the choice of either scheme is \emph{not}
in fact immaterial for issues that are widely regarded as being highly relevant to modern particle physics.
\end{itemize}
It should once more be stressed that the above account corresponds to the ``old'' theory of renormalization, as
initially developped mainly on the basis of QED. As will be seen in section \ref{SMtriv} however, even though the
modern view on renormalization that became the orthodoxy in the 1970s differs substantially from this account,
many of the above introduced concepts and ideas still figure prominently also within the modern view.

\subsection{The Renormalized Self-Interacting Scalar Field}\label{phi4renorm}

\noindent It is instructive to exemplify some elements of the slightly abstract account given in the previous subsection
in terms of a concrete, simple model.
In ordinary (i.e, four-dimensional) spacetime, the simplest, physically relevant quantum field theory is that of
a real scalar field, $\phi$, governed by a quartic self-interaction, i.e., as described by a density of the form
\begin{equation}\label{phi4Lagr}
\mathscr{L} \; := \; \mathscr{L}(m^2,\lambda,\phi) \; = \; - \frac{1}{2}(\partial \phi)^2 \: - \: \frac{1}{2}m^2 \phi^2 \: - \: \frac{1}{4!} \lambda \phi^4 \qquad \quad m^2 > 0 \qquad \lambda > 0
\end{equation}
The theory based on (\ref{phi4Lagr}) (often colloquially referred to as $\lambda \phi^4$-theory) is essentially nothing
but the part of the standard model exclusively involving the Higgs field\footnote{The (minimal) standard model Higgs field is taken to transform
according to the fundamental representations of weak isospin and weak hypercharge (i.e., it is represented as a doublet 
of $\mbox{SU}(2)$ weak isospin and as a singlet of $\mbox{U}(1)$ weak hypercharge, respectively),
but apart from these subtleties, it is governed by essentially the density (\ref{phi4Lagr}).
In four dimensions, this density defines the only nontrivial renormalizable model for a scalar field with positive definite energy.}  and is therefore 
also highly relevant for the specific arguments put forward in section \ref{naturalness}.\\
As is well known, for $\lambda \phi^4$-theory, the proper Green's functions referred to in item (ii) in subsection
\ref{renormalization}, correspond to $n=2$ (i.e., the ``propagator'') and $n=4$.
The removal of ambiguities from these two functions is achieved through renormalizations of the parameters $m^2$ and $\lambda$
and the field variable $\phi$ in (\ref{phi4Lagr}) and, according to the previous discussion, is equivalent to the removal
of ambiguities from all Green's functions\footnote{Although the renormalizations of $m^2$, $\phi$, on the one hand, and
$\lambda$ on the other hand are related essentially to the divergences in the two- and four-point functions, respectively,
these renormalizations are not independent (i.e., the perturbative expressions of these two functions ``contain'' each other
in a sense and in order to remove the ambiguities from one of the two functions at any given order in perturbation theory, it is
necessary that the ambiguities from the other function have already been removed to the previous order - this is one
of the nontrivial aspects of renormalization theory referred to earlier).}.
A formal expansion of the (Fourier-transformed) propagator according to (\ref{GLformula}) gives rise to the following expression
\begin{equation}\label{phi4-2point2}
\widetilde{G}^{(2)}(p) \; = \; \widetilde{G}_F(p) \: + \: \widetilde{G}_F(p)\frac{\Sigma(p)}{i}\widetilde{G}_F(p) \: + \: \widetilde{G}_F(p)\frac{\Sigma(p)}{i}\widetilde{G}_F(p) \frac{\Sigma(p)}{i}\widetilde{G}_F(p)  \: + \: \cdots 
\end{equation}
where $\widetilde{G}_F(p) \; := \; (p^2 + m^2)^{-1}$ denotes the free-field propagator and where the proper two-point 
function, $\Sigma(p)/i$ (also referred to as the ``self-energy''), contains divergent momentum ``loops'' to all orders 
in perturbation theory\footnote{The subscript ``F'' in the symbol for the free-field propagator refers to the implicitly understood Feynman
(i.e., $-i\epsilon$) prescription for dealing with the poles (in fact, the prescription is implicitly understood also in expressions such
as eq. (\ref{phi4-2point5})).}.
Note however that the geometric series contained in (\ref{phi4-2point2}) can formally be summed to yield
\begin{equation}\label{phi4-2point5}
\widetilde{G}^{(2)}(p) \; = \; \frac{1}{p^2 + m^2 + i \Sigma(p)}
\end{equation}
(with the understanding that (\ref{phi4-2point5}) is indeed a purely formal expression which is \emph{defined} by
the, also formal, series (\ref{phi4-2point2})).
The expression (\ref{phi4-2point5}) gives some rudimentary insight into the renormalization of mass in $\lambda \phi^4$-theory.
Indeed, since $m^2$ is just an arbitrary, i.e., ``bare'', parameter of the theory, one is free to simply define
a renormalized mass, $m_{r}$, according to
\begin{equation}\label{renormcond}
m_r^2 \; := \; m^2 \: + \: \left. i \Sigma(p) \right|_{p^2 = - M^2} 
\end{equation}
where $M$ denotes a certain specified mass scale (a so-called subtraction point)\footnote{Usually, $M^2$ is simply
chosen equal to $m_r^2$, so that the renormalized mass is only implicitly defined by (\ref{renormcond}) and equals
the actual ``physical'' mass, i.e., the mass corresponding to the poles of the propagator (\ref{phi4-2point5}) (at least formally).
Eq. (\ref{renormcond}) represents a so-called renormalization condition (or prescription) and it is customary to impose the further condition
\begin{equation}\label{renormcond3}
\left. \frac{d \Sigma (p)}{dp^2} \right|_{p^2 = - m_r^2} \; = \; 0
\end{equation}
The standard renormalization conditions then simply express that the formal propagator (\ref{phi4-2point5}) 
has a simple pole at $p^2 = - m_r^2$ of residue $1$ (as noted, expressions are usually first regularized, so that
eqs. (\ref{renormcond}), (\ref{renormcond3}) contain only strictly finite quantities at each perturbative order;
but even in that case, (\ref{phi4-2point5}) still represents a mere formal series, as defined by eq. (\ref{GLformula})).}.
A more systematic approach towards the renormalization of $\lambda \phi^4$-theory proceeds via the method of
counterterms. To this end, it is useful to introduce a slight change in notation and denote
the bare parameters and field in (\ref{phi4Lagr}) with a subscript ``0'' and to then express these
bare quantities in terms of their renormalized versions, denoted by the same symbols, but without a subscript,
and three freely adjustable parameters, $\delta_m$, $\delta_{\lambda}$ and $Z$, as follows
\begin{equation}\label{counterterm1}
\phi_0 \; := \; \sqrt{Z} \phi \qquad \quad m^2_0 Z := \delta_m \: + \: m^2 \qquad \quad \lambda_0 Z^2 \; := \; \delta_{\lambda}  \: + \: \lambda
\end{equation}
(note that, apart from their perturbative character, eqs. (\ref{renormcond}), (\ref{counterterm1}) are again in complete
analogy with the renormalization of mass in classical electrodynamics according to (\ref{CEDmassrenorm})).
Substitution of (\ref{counterterm1}) into (\ref{phi4Lagr}) leads to the following decomposition
\begin{equation}
\mathscr{L}(m_0^2,\lambda_0,\phi_0) \; = \; \mathscr{L}(m^2,\lambda,\phi) \: + \: \mathscr{L}(\delta_m\delta_Z^{-1}, \delta_{\lambda} \delta_Z^{-2},\sqrt{\delta_Z}\phi) 
\end{equation}
with $\delta_Z := Z - 1$ and the coupling constants $\delta_m, \delta_{\lambda}, \delta_Z$ in the second term of
this decomposition, i.e., exclusively consisting of the ``counterterms'', are to be specifically adjusted so as to
cancel the divergent contributions to $\widetilde{G}_{\mbox{\scriptsize 1PI}}^{(2)}$ and $\widetilde{G}_{\mbox{\scriptsize 1PI}}^{(4)}$ at each order in perturbation theory.
Mathematically, a very convenient way to effectuate such cancellations is in terms of dimensional regularization.
In essence, this boils down to expanding the bare parameters, $m_0$, $\lambda_0$ and $Z$ as formal power series in
the inverse of the regulator, $\epsilon$, from which the dimension of spacetime is taken to temporarily deviate from
$4$ and to then express each coefficient in this expansion as a power series in the (yet to be) renormalized coupling $\lambda$.
Thus, for the coupling itself, for instance, this means that
\begin{equation}\label{expansionbare}
\lambda_0   \; := \; \mu^{\epsilon} \biggl( \lambda \: + \: \sum_{\nu = 1}^{\infty} \frac{a_{\nu}(\lambda)}{\epsilon^{\nu}} \biggr)
\end{equation}
with the understanding that the coefficients $a_{\nu}$ have power series expansions of the specific form 
$a_{\nu}(\lambda) = \sum_{j = \nu}^{\infty} a_{\nu j} \lambda^j$, $a_{\nu j} \in \real{}$ and with $\mu$
denoting an arbitrary parameter with the dimension of mass\footnote{This parameter is introduced for the sole purpose
of keeping the renormalized coupling constant, $\lambda$, dimensionless - and hence power series expansions in this 
constant meaningful - in $d := 4 - \epsilon$ dimensions.
Also, the specific form of the coefficients $a_{\nu}$ is motivated by the fact that diagrams with $n$ closed
loops - which essentially diverge as $\epsilon^{-n}$ - do not occur until $n$th order in perturbation theory.}.
The coefficients $a_{\nu j}$ in the expansion (\ref{expansionbare}) and its analogues for $m_0^2$ and $Z$ can then
be specifically adjusted to cancel the parts of the loop integrals that diverge upon removing the regulator, i.e.,
upon taking the limit $\epsilon \rightarrow 0$.
However, a proper discussion of this is beyond the scope of the present treatment\footnote{For a technically more detailed treatment,
see \citeN{Holman2} and references therein.} and for future reference I will simply state the explicit form of (\ref{expansionbare})
that leads to the required cancellations to order $\lambda^2$
\begin{equation}\label{bareseriesexp1}
\lambda_0  \; = \;  \mu^{\epsilon} \Bigl( \lambda \: + \: \frac{3}{16 \pi^2} \frac{\lambda^2}{\epsilon} \Bigr) 
\end{equation}
Even upon discarding any qualms about the mathematical consistency of the renormalization procedure as just sketched
explicitly for $\lambda \phi^4$-theory (recall footnotes \ref{fieldsubtleties}, \ref{perturbdiff}), a conceptually unsettling feature about
this procedure, that was somewhat glossed over in the previous, is the following.
The definition (\ref{expansionbare}) evidently restricts the arbitrariness in the bare coupling $\lambda_0$ to a 
\emph{particular form}. That is, it corresponds to a particular subtraction scheme, which is just the minimal
scheme described earlier\footnote{Note that (\ref{expansionbare}) corresponds to essentially a Laurent series 
subtraction with nonzero coefficients only for the poles in $\epsilon$ : had (\ref{expansionbare}) also contained
positive powers of $\epsilon$ this would have generated infinitely many additional, freely adjustable parameters
in the finite parts of Feynman diagrams. It is in this clear sense that the subtraction defined by (\ref{expansionbare})
is indeed minimal.}.
As noted before however, in contrast to the situation for (nonlinear observables of) free fields, there exists
no canonical subtraction prescription for interacting fields. It is in fact clear that infinitely many different
subtraction choices are possible and that, as a consequence of this, there is an essential ambiguity in all
renormalized expressions (in the present context, this is in fact also true because of the presence
of the arbitrary mass scale, $\mu$). This seems irreconcilable with the extra-ordinary empirical adequacy
claimed for theories such as QED.
On the other hand, even though renormalized parameters, such as masses and coupling constants, are not actually
calculable by theory, they can be determined by experiment\footnote{See also \citeANP{Schwinger11} \citeyear{Schwinger11,Schwinger12}.}.
Once this is done for a particular subtraction scheme, value of $\mu$ and degree of accuracy, no arbitrariness is left
and calculable effects should be retrievable from the theory\footnote{For the specific MS scheme (\ref{expansionbare}),
renormalized parameters are defined up to finite transformations that depend only on the arbitrary mass scale $\mu$.
This may be expressed by saying that the MS scheme corresponds to the specification of a one-parameter family of 
renormalization conditions in the sense of eqs. (\ref{renormcond})-(\ref{renormcond3}), except that the physical 
significance of these conditions is not as clear in this case, because $- \mu^2$ is not necessarily equal to the 
(squared) value of some external momentum.}.\\
However, how should a particular value of $\mu$ be determined, or, more specifically, what is the physical significance
of this parameter ?
At a general conceptual level, the presence of an arbitrary mass scale in the renormalized theory seems quite awkward.
In fact, before renormalization, the theory contained two parameters, $m_0$ and $\lambda_0$, and the scale $\mu$ entered 
only as part of the regularization scheme.
Shouldn't this scale therefore disappear, as a matter of principle, after removing the regulator~?
Furthermore, how to justify the appearance of $\mu$ when using an explicit cutoff scheme, in which 
coupling constants do not become dimensionful at intermediate stages~? Although the answers to these questions are 
far from obvious, according to the received view, renormalization of \emph{any} renormalizable quantum field theory 
is unavoidably accompanied by the appearance of an arbitrary mass scale into the theory.
This is the so-called ``sliding renormalization scale'' of Gell-Mann and Low, which is central to the
renormalization group concept discussed further in section \ref{SMtriv}.
The appearance of such a scale is often referred to as the \emph{anomalous breakdown of classical scale invariance}
(or simply as the scale or conformal or trace anomaly), the idea being that, discarding possible bare mass parameters, any strictly renormalizable
field theory should be classically scale invariant, as it can contain no dimensionful parameters\footnote{The seemingly unavoidable appearance of a dimensionful 
parameter as a result of the renormalization process is sometimes also referred to as \emph{dimensional transmutation}.
The latter concept is more subtle however and refers to situations in which, as a result of including radiative corrections,
a dimensionless, free parameter (i.e. a coupling constant) is actually ``traded in'' for another free parameter which
is \emph{dimensionful}, namely the sliding renormalization scale.
This happens for instance in certain models displaying spontaneous symmetry breaking (where the sliding scale actually
corresponds to the expectation value of an external field, rather than a certain point in momentum space; see 
\citeN{ColWei}) and in asymptotically free quantum field theories, such as QCD.\label{dimtrans}}.

\section{Expected Triviality of the Standard Model}\label{SMtriv}

\noindent As is well known, initially in the development of quantum field theory, renormalizability was taken to correspond
to a fundamental principle of some sorts, necessary for guiding theory selection.
In fact, this view - which was clearly central to the discussion of renormalization in the previous section - continued 
to be widespread well into the 1970s, during the construction of the standard model of particle physics, which is strongly
suggested as the simplest, concrete quantum field theoretical model meeting (i) a few basic empirical facts 
(such as the chiral asymmetry exhibited in particle interactions, the \emph{global} $\mbox{SU}(2)$ weak isospin and 
$\mbox{SU}(3)$ colour symmetries), which partially determine the menu of relevant fields, (ii) the so-called
gauge principle and (iii) the requirement of (perturbative) renormalizability\footnote{See e.g. \citeN{Weinberg3}.
Contrary to what is claimed in some references, for theories involving spin-one fields, renormalizability, by itself, 
does not imply gauge invariance. In fact, even for such theories, the conditions (ii) and (iii) are quite independent, 
in that neither condition implies the other (this is probably best illustrated by the existence of certain models with explicit
mass terms for the Yang-Mills fields, such as the St\"uckelberg-type models - which are gauge invariant but not 
renormalizable - and the Curci-Ferrari model - which is renormalizable, but not gauge invariant). 
Although it \emph{does} appear impossible to formulate a consistent theory of spin-one fields that is both
renormalizable by power counting and perturbatively unitary \emph{unless} it is a (spontaneously broken) gauge theory
of Yang-Mills type (this is at least the received view), it is rather doubtful that this provides a compelling case
for gauge invariance, given the changed status of perturbative arguments and renormalizability (as the discussion in the present section should make clear).}.
During that same period  however, the traditional view of renormalization began to be challenged.
The perspective that finally emerged - and in which renormalizability no longer had a fundamental status - 
was an amalgamation of several semi-independent developments, which centred around a then newly reached 
(re-)understanding of two key concepts belonging to separate disciplines of physics, namely 
\begin{itemize}
\item[(i)] the notion of scale dependence of physical parameters within the context of quantum field theory, and
\item[(ii)] the notion of critical phenomena within the context of statistical physics\footnote{For a more elaborate
discussion of the historical developments with a slightly different emphasis, see \citeN{CaoSch}.}.
\end{itemize}
The idea that physical parameters, such as the charge and mass of the electron, might be scale dependent first emerged in
the early 1950s and was soon developped further in terms of the notion of a sliding renormalization scale (or subtraction
point), as introduced by \citeANP{GelLow} in their famous 1954 paper on the short-distance behaviour of QED, which
in turn lies at the heart of the renormalization group concept.
For a number of reasons however, the crucial insights on renormalization obtained in these years were not
fully absorbed for a long time.
With the discovery of anomalies in the early 1970s, the key lesson of the Gell-Mann and Low approach - which was the
realization that a naive dimensional analysis breaks down precisely because of renormalization - was essentially rederived
afresh and subsequently became a central component of the modern view on quantum field theories\footnote{For a detailed discussion 
of the Gell-Mann and Low paper, both with regards to its actual contents and its historical significance, see \citeN{Weinberg2}. 
In addition to this key paper, essentially simultaneous investigations into the scale dependence of physical parameters 
were undertaken by a number of authors, including Dyson, St\"uckelberg and Peterman and, as will be recalled shortly, Landau.}.\\

\subsection{Perspectives on Quantum Field Theories : Landau, Gell-Mann and Low, Wilson}\label{QFTperspectives}

\noindent At around the same time that Gell-Mann and Low presented their version of the renormalization group, an independent
scaling analysis of the short-distance structure of QED, which in fact exposed a very serious potential difficulty 
regarding the internal consistency of the theory, was put forward by \citeN{Landau} and collaborators.
In modern language, Landau's famous ``zero charge'' argument boils down to the statement that the so-called beta-function 
of QED (see below), in terms of the electric charge, $e_{\mu}$, viewed as a function of the energy scale, $\mu$, ``under 
consideration'', is given by
\begin{equation}\label{QEDbetafunction}
\mu \frac{d}{d \mu} e_{\mu} \; := \; \beta (e_{\mu}) \; = \; \frac{1}{12\pi^2} e_{\mu}^3 \: + \: \mathcal{O}(e_{\mu}^5)
\end{equation}
Upon neglecting all higher order contributions, eq. (\ref{QEDbetafunction}) integrates to (expressed in terms of the fine 
structure constant, $\alpha_{\mu} = e_{\mu}^2 / 4\pi$, rather than the charge)
\begin{equation}\label{QEDRG}
\alpha_{\mu} \; = \; \frac{\bar{\alpha}}{1 - \frac{2 \bar{\alpha}}{3 \pi} \log ( \mu / \bar{\mu})}  \qquad \qquad \bar{\alpha} := \alpha_{\bar{\mu}} 
\end{equation}
for some arbitrary mass scale $\bar{\mu}$.
Thus, if eq. (\ref{QEDRG}) continues to correctly represent the behaviour of
the fine structure constant at high energies, QED is not a consistent theory,
since $\alpha_{\mu}$ blows up at some very large, but finite energy scale, 
$\mu_{\infty}$ (the so-called ``Landau pole'')
\begin{equation}\label{Landaughost}
\mu_{\infty} \; = \; \bar{\mu} e^{3 \pi / 2 \bar{\alpha}} \; \simeq \; 10^{276} \, \mbox{GeV}
\end{equation}
(where $\bar{\mu}$ and $\bar{\alpha}$ were equated with the ``usual'' electron mass $\simeq 0.5 \mbox{MeV}$
and fine structure constant $\simeq 1/137$, respectively - i.e., as defined at sufficiently macroscopic distances; cf. also subsection \ref{EFTparadigm}). 
On the other hand, $\bar{\mu}$ is an arbitrary integration limit and eq. (\ref{QEDRG}) is valid equally well for $\mu < \bar{\mu}$.
But it is clear that if $\bar{\mu}$ is taken to infinity for \emph{any} fixed, finite value of $\mu$, $\alpha_{\mu}$ 
necessarily vanishes.
In other words, by sending $\bar{\mu}$ to infinity, the Landau pole is removed and QED becomes defined
on all distance scales, but only as a trivial theory\footnote{A similar verdict about quantum field theory in general
was reached by \citeN{Haag2}, within the mathematically rigorous context of the Wightman formulation for quantum field theories.
Haag's theorem essentially states that representations of the canonical commutation relations for quantum fields of different 
interaction strengths are unitarily inequivalent. This means in particular that the interaction picture does not exist, 
strictly speaking, so that the starting point (\ref{GLformula}) for the entire perturbative approach is mathematically ill-founded.
Although no renormalization ever enters the discussion here, it seems rather clear that the common element in both 
triviality arguments is the ambiguity caused by the attempted time ordering of Wick polynomials in (\ref{GLformula}).}.
Now the value (\ref{Landaughost}) is of course absurdly large (far larger, in fact, than the total mass of the 
observable universe), but this doesn't change the fact that the Landau pole poses a serious threat to the
internal theoretical consistency of QED (or any other theory with a similar short-distance behaviour; see also subsection \ref{SMRGflow}).
Moreover, as Landau also pointed out, if $\mathscr{N}$ species of spin-$\frac{1}{2}$ particles exist with the same 
charge as the electron, the pole is effectively lowered to $10^{276/\mathscr{N}} \, \mbox{GeV}$.
Nevertheless, it is clear that eq. (\ref{QEDRG}) does not take into account higher-order corrections (which clearly become 
increasingly important at high energies), so that, in the form in which it has been presented here, the argument
is not in fact conclusive. 
This realization is also implicit in the work of \citeANP{GelLow}, who presented a case on essentially the same basis 
for the breakdown of perturbative QED at very high energies, but at the same time acknowledged the possible existence 
of a nontrivial, finite UV ``fixed point'' of the renormalization group equations.

\subsubsection*{Conformal Anomaly and Possible Short-Distance Scenarios}

\noindent The starting point of the \citeANP{GelLow} analysis is the observation that prima facie expectations about
scale invariance in QED are violated because of renormalization effects.
Indeed, according to naive dimensional analysis, a Coulomb potential, having the dimension of mass, should scale as
$1/r$, but this is not in fact the case : although the leading order term behaves this way, first-order radiative corrections
involve terms with a $\log (1/r)$ dependence.
The source of this anomaly, as \citeANP{GelLow} demonstrate, ultimately has to be sought in the necessity of charge 
renormalization, together with the standard prescription to \emph{define} electric charge in terms of the large-distance
behaviour of the Coulomb potential.
This suggests that, by defining electric charge instead in terms of a renormalization prescription at some arbitrary
distance, $R$, scale invariance might be restored for $R$ small enough.
Thus enters the sliding renormalization scale.
\citeANP{GelLow} then continue to investigate the consequences of this important idea and conclude by noting that
there are essentially two possibilities for the ultra-short distance limit.
Using the representation for the renormalized charge as a function of the sliding subtraction point 
(expressed in terms of an energy scale, $\mu$, rather than a distance scale) given by eq. (\ref{QEDbetafunction}), 
one sees that indeed either
\begin{itemize}
\item[(i)] $e_{\mu}$ does not have a finite limit for $\mu \rightarrow \infty$, in which case the
``bare charge'' is infinite and the theory is very probably inconsistent (because of the presumable presence of a Landau 
pole), or
\item[(ii)] $e_{\mu}$ converges to a finite limit, $e_{*}$, for $\mu \rightarrow \infty$.
In this case, $e_*$ is a fixed point of the so-called Gell-Mann and Low function $\psi$ (and a zero of the
beta-function (\ref{QEDbetafunction})) and scale invariance is essentially recovered asymptotically
\end{itemize}
It is important to note however, that if a fixed point exists, it is not trivial.
Indeed, if $e_{\mu}$ is sufficiently small for some particular value of $\mu$, perturbation theory should be valid
and eq. (\ref{QEDbetafunction}) then shows that $e_{\mu}$ \emph{increases} with growing $\mu$.
Put slightly differently and independent of whether or not a fixed point actually exists, the Gell-Mann and Low analysis shows
that \emph{perturbation theory inevitably breaks down at ultrahigh energies}.
Since all known renormalization techniques are intrinsically perturbative, this means that even if a particular theory
possesses a stable, nontrivial ultraviolet fixed point, it cannot be renormalizable in the conventional sense.
Nevertheless, any such theory is still expected to be internally consistent because of its seemingly reasonable high-energy behaviour.
This thus suggests replacing the condition of renormalizability as a criterion for maintaining theoretical consistency,
by the more general condition that a stable fixed point exists in the far-ultraviolet regime.
Following Weinberg, theories that satisfy this more general condition will be referred to as \emph{asymptotically safe}\footnote{Cf. \citeANP{Weinberg4} \citeyear{Weinberg4,Weinberg5}.
An example of such a theory is an interacting scalar field theory formulated in a five-dimensional flat spacetime.
If the Hamiltonian is required to be symmetric under reflections, $\phi \rightarrow - \phi$ (so that it is bounded 
from below), the theory cannot be renormalizable, since the only renormalizable scalar field interaction in five spacetime 
dimensions is a cubic term.
Nevertheless, there appears to be evidence for the existence of a nontrivial ultraviolet fixed point in this case. 
See \citeN{Weinberg4}.}.\\
As already noted, the importance of the Gell-Mann and Low results was not immediately recognized as such and
the whole idea of physical parameters being scale dependent, as precisely governed by fundamental equations, was essentially 
rediscovered in the early 1970s. In what follows, the discussion will be based on these more recent developments.
Obviously, any interacting $n$-point function (\ref{GLformula}) is defined in terms of only bare parameters and fields
and is thus independent of any regularization parameter, such as the arbitrary mass scale $\mu$ 
that was seen to enter the dimensional regularization scheme.
Thus, upon making use of the elementary relation $\Phi_0 = \sqrt{Z_{\Phi}} \Phi$, for a generic bare field, $\Phi_0$,
its renormalized version, $\Phi$, and the corresponding wavefunction renormalization constant $Z_{\Phi}$ (cf. the first
expression in (\ref{counterterm1}), in the particular case of $\lambda \phi^4$-theory), differentiation of a bare
$n$-point function with respect to $\mu$ gives 
\begin{eqnarray}\label{CalSymeq}
\mu \frac{d}{d \mu} G_0^{(n)} & = & \mu \frac{d}{d \mu} \left( Z_{\Phi}^{n/2}(m, g, \mu) G^{(n)} (m , g , \mu ) \right)     \nonumber \\
                            & = & Z_{\Phi}^{n/2} \left( \mu \frac{\partial}{\partial \mu} \: + \: \beta (g) \frac{\partial}{\partial g} \: + \: m \xi(g) \frac{\partial}{\partial m} \: + \: \frac{n}{2} \gamma (g) \right) G^{(n)}  \; = \; 0
\end{eqnarray}
where for definiteness the case of only one (dimensionless) coupling constant and one mass, denoted by respectively $g$ 
and $m$ was taken, with $m$ moreover such that $m \ll \mu$, and where the functions $\beta$, $\xi$  and $\gamma$ are defined according to\footnote{The fact 
that these functions (effectively) only depend on $g$ follows from the assumed properties. That is, eq. (\ref{CalSymeq}), together with
the fact that $g$ is dimensionless imply that the dimensionless function $\beta := \mu dg / d \mu$, for instance, a priori 
is of the form
\begin{equation}
\beta \; := \; \beta (g , m / \mu)
\end{equation}
Since, by assumption, $m \ll \mu$ for the entire range of relevant values of $\mu$, this implies that effectively 
$\beta = \beta (g, 0) =: \beta (g)$, (strictly speaking, this requires also that there are no zero-mass singularities;
see \citeN{Weinberg6}, section 18.2, for a discussion of this last point).
Entirely analogous remarks apply to the functions $\xi$ and $\gamma$.}
\begin{equation}\label{betaxigamma}
\beta(g) \; := \; \mu \frac{d}{d \mu}g   \qquad \quad \xi(g)  \; := \; \frac{\mu}{m} \frac{d}{d \mu} m \qquad \quad \gamma ( g ) \; := \; \mu \frac{d}{d \mu} \ln Z_{\Phi}  
\end{equation}
Eq. (\ref{CalSymeq}) is usually referred to as the \emph{Callan-Symanzik} or \emph{renormalization group equation}\footnote{\citeN{Callan}, \citeN{Symanzik}. 
It should be noted that some authors reserve this terminology merely for the definition of the beta-function, i.e. the first
equation in (\ref{betaxigamma}). It should also be noted that the notation in eqs. (\ref{CalSymeq})-(\ref{betaxigamma}) 
appears to be standard, in the sense that these equations are independent of any particular regularization scheme.
In particular, $\mu$ just represents the scale that unavoidably enters \emph{any} renormalized theory, according the
modern view of things.}.
The function $\beta$ governs the scale dependence of the coupling constant, $g$, and it seems quite clear that a theory for 
which this function is nonzero cannot be scale invariant, essentially because $g$ is no longer spacetime independent.
In fact, it is not difficult to show that the breakdown of any classical scale symmetry directly manifests itself in the (vacuum expectation value of the)
trace of the  (possibly ``improved'') stress-energy tensor acquiring a nonzero value, which, moreover, is directly proportional to the beta-function.\\
An explicit expression for $\beta$ in perturbation theory is most conveniently derived in terms of dimensional 
regularization and the minimal subtraction scheme briefly outlined in the previous section\footnote{Apart from the beta function,
an important role in renormalization group analyses is also attributed to the function $\gamma$, the so-called 
\emph{anomalous dimension}. The idea is roughly that upon treating field dimensions as new, dynamical degrees of
freedom dependent upon the function $\gamma$, effective scale invariance can explicitly be recovered at short distances
if the theory in question has an ultraviolet fixed point (this is most easily seen from rewriting eq. (\ref{CalSymeq}) 
in the form $\mu d G^{(n)} / d \mu = - n \gamma G^{(n)}/2$, which, for $n=2$, tentatively integrates to the following form of the propagator
\begin{equation}
G^{(2)}(p^2) \sim \frac{1}{p^{2 - \gamma_*}}   \qquad \quad \gamma_* \: :=  \: \gamma (g_*)
\end{equation}
for large (Euclidean) momenta, $p$, assuming a nontrivial UV fixed point $g_*$ indeed exists).
Historically, the general idea of absorbing interaction effects into field dimensions was first introduced by Wilson
within the context of his operator product expansion (OPE) framework, as applied to the strong nuclear interaction.
Although in that case, the idea turned out not to work (see the main text), it did prove to be important in applications
of the renormalization group to critical phenomena, where the anomalous field dimensions translate into the nontrivial
(i.e., non-Landau related) parts of the critical exponents.}.
Although the identity (\ref{expansionbare}) appeared in the specific context of $\lambda \phi^4$-theory, it is not difficult
to see that a completely analogous definition can be given for a generic dimensionless coupling $g$. Writing such an identity
symbolically as $g_0 = \mu^{\epsilon} \bar{g} (g, \epsilon)$, 
differentiation with respect to $\mu$ and multiplying the result with $\mu$ gives
\begin{equation}
0 \; = \; \mu \frac{d}{d \mu} g_0 \; = \; \mu^{\epsilon} \biggl( \epsilon \bar{g} \: + \: \mu \frac{\partial \bar{g}}{\partial g} \frac{d g}{d \mu} \biggr)
\end{equation}
so that
\begin{equation}
\beta_{\epsilon} (g) \; = \; - \frac{\epsilon \, g_0}{ \partial g_0 / \partial g}
\end{equation}
Substitution of an expansion of the form (\ref{expansionbare}) for the bare coupling into this equation then yields
\begin{equation}\label{betafnctitoa1}
\beta_{\epsilon} (g) \; = \; - \epsilon g \: - \: \left( 1 \: - \: g \frac{\partial}{\partial g} \right) a_1 (g) \: + \: \mathcal{O} (1 / \epsilon)
\end{equation}
If the theory is renormalizable (or, more generally, if it is asymptotically safe), it should be possible to remove the
regulator without generating infinities. In particular, all terms of order $1/\epsilon$ and higher are necessarily absent 
in that case and $\beta (g) := \beta_0 (g)$ should be a finite quantity, which, according to (\ref{betafnctitoa1}), is 
determined entirely by the coefficient, $a_1$, of the lowest order pole in an expansion of the form (\ref{expansionbare}) 
of the bare coupling in terms of the renormalized coupling\footnote{For a more sophisticated argument leading to the
same conclusion, see \citeN{Weinberg6}, section 18.6.}.
Disregarding theories for which $\beta$ vanishes identically, such as $N=4$ supersymmetric $\mbox{YM}_4$-theory, 
of which the physical relevance is unclear at the time of writing\footnote{The reason why a four-dimensional Yang-Mills
theory with a $N=4$ supersymmetry is nevertheless considered to have potential physical significance by some physicists
is because it for instance appears to explicitly realize Montonen-Olive duality, which relates different (i.e., perturbative
and non-perturbative) parametrizations of the theory.
It has also been conjectured to be dual to a so-called type IIB string theory on $S^5 \times \mbox{AdS}_5$ 
(cf. \citeN{Maldacena}; here $\mbox{AdS}_5$ denotes five-dimensional anti-de Sitter space).\label{N4SUSYYMnote}}, there are essentially 
two main possible types of behaviour for this function, which are just the appropriate analogues of the QED cases (i) 
and (ii), already specified by Gell-Mann and Low.
That is, in the far-ultraviolet regime, either
\begin{itemize}
\item[(i)$'$] $g_{\mu}$ does not have a finite limit as $\mu \rightarrow \infty$, in which case again the theory is very 
probably inconsistent, or
\item[(ii)$'$] $g_{\mu}$ converges to a finite limit, $g_{*}$, as $\mu \rightarrow \infty$, an ultraviolet (stable) fixed 
point of the renormalization group flow, in which case $\beta (g_{*}) = 0$, and the theory is said to be asymptotically safe.
\end{itemize}
In the second case however, it is now necessary to further distinguish between \emph{two} fundamental scenarios, of which only 
one was considered by Gell-Mann and Low, as it is the only possibility in the specific case of QED.
That is, if $g_{\mu} > 0$ converges to a finite limit, $g_*$, at very high energies, either
\begin{itemize}
\item[(iia)$'$] $g_{*} > 0$, which is possible only if $\beta(g) \geq 0$ on the interval $0 \leq g \leq g_*$, with $\beta(0)=0$, or
\item[(iib)$'$] $g_{*} = 0$, which is possible only if $\beta(g) \leq 0$ on some interval $0 \leq g \leq g_{\bar{\mu}}$
\end{itemize}
In making these statements, strictly speaking, it is necessary to assume that perturbation theory is valid at \emph{some}
point. But, as this is a premise implicit in almost every treatment of quantum field theory relevant to high-energy
physics (and certainly underlies all of the discussion of quantum field theory in the present article), this does
not amount to any serious loss of generality - at least as far as the present context is concerned.
In the perturbative regime then, the beta function, to lowest order, can be expressed as
$\beta(g) = C g^n$, for some positive integer power $n$ and real, nonzero constant $C$ (in fact, within a dimensional regularization
scheme, by eq. (\ref{betafnctitoa1}), $C$ equals $(n-1)a_{1n}$, where $a_{1n}$ denotes the first non-zero coefficient in the power series expansion of $a_1(g)$ for $n>1$).
For $C<0$, the existence of an ultraviolet fixed point, $g_*=0$, is then implied\footnote{Intuitively, this is plausible,
since $g$ decreases (causing the lowest-order approximation to become even more accurate) with increasing energy, if $C<0$.
Explicit integration (completely analogous to the QED case, with $C = 2/ 3 \pi$ and $n=2$) gives
\begin{equation}
g_{\mu} \; = \; \left( \frac{1}{\bar{g}^{1-n} - (n-1) C \log (\mu / \bar{\mu})} \right)^{1/(n-1)} \qquad \bar{g} := g_{\bar{\mu}} \quad n >1
\end{equation}
Clearly, for $C<0$, $g_{\mu} \rightarrow 0$ in the ultra-high energy limit, which is just asymptotic freedom.
Incidentally, if again only the leading order term is taken into consideration, the Landau pole re-appears at 
$\mu_{\infty} = \bar{\mu} \exp \left( \frac{1}{(n-1)C\bar{g}^{n-1}} \right) < \bar{\mu}$ and is thus shifted to the infrared region.}.
On the other hand, for $C>0$, a nontrivial fixed point, $g_* > 0$ - a so-called \emph{Wilson-Fisher fixed point}\footnote{Although by 
definition, a Wilson-Fisher fixed point occurs at nonzero coupling - i.e., is nontrivial - it need not correspond to 
the short-distance limit of the theory. Indeed, the first nontrivial fixed point for which evidence was found
occurs in an Ising-type model formulated in $d<4$ spacetime dimensions and is an \emph{infrared} fixed point. Cf. \citeN{WilFis}.
A trivial fixed point (which in this particular case occurs for $d \geq 4$), is also referred to as a \emph{Gaussian}
fixed point, as the critical behaviour for this case is indistinguishable from that of a free massless field theory
(which in the particular case of a scalar field is also referred to as a Gaussian model).
Finally, it should be noted that there is a certain circularity in the present discussion of fixed points, since the 
argument for the particular form of the beta-function - i.e.,  involving only $a_1$ (cf. eq. (\ref{betafnctitoa1})) - 
was seen to depend on the assumption of asymptotic safeness, that is, the assumption that an ultraviolet stable fixed point actually exists.\label{WiFiFP}} -
may exist as $\mu$ goes to infinity, but it cannot be reached perturbatively, whereas $g_0 := \lim_{\mu \rightarrow 0}g_{\mu} = 0$ 
is an infrared (stable) fixed point, since $\beta(0)=0$.\\
Scenario $\mbox{(iib)}'$ corresponds to an entirely new type of behaviour, the possibility of which was not recognized until 
the early 1970s, when the short-distance behaviour of generic Yang-Mills gauge theories became understood.
Theories in this class, which become non-interacting field theories at asymptotically large energies, are called
\emph{asymptotically free}. Such theories can be regarded as perturbatively renormalizable (or at least this is usually
claimed), since the ultraviolet fixed point associated with any one of them is Gaussian.
It is because of their good ultraviolet properties that theories for which $\beta < 0$ are sometimes attributed
a nearly divine status. As will become clear however, such a view is a bit premature.

\subsubsection*{Scales in Physics : Wilson's View of the Renormalization Group}\label{WilsonviewRG1}

\noindent A somewhat different approach towards the issue of the short-distance structure of generic quantum field
theories was pioneered in a series of articles by Wilson, approximately in the period 1965-1972, and was largely
inspired by the wish to improve upon the (generalized) Landau description of critical phenomena.
In his efforts to do so, Wilson adopted a radically novel perspective on how to assess the importance 
of fluctuations in critical phenomena and, as a result, managed to lay the foundations for a new paradigm for quantum field theory (cf. subsection \ref{EFTparadigm}).
The fundamental point of departure in this approach is to take seriously the idea that the order parameter
is a field from which microscopic fluctuations have already been averaged out\footnote{It is recalled that 
Landau's (generalized) theory of phase transitions is fundamentally based on the notion of an \emph{order parameter}, 
$\varphi$, which is in general a dynamic, fluctuating object, for which the thermal average, $\langle \varphi \rangle$, vanishes 
in the unordered (i.e., symmetric and almost invariably high-temperature) phase, but is nonzero in the ordered (i.e., broken
symmetry) phase. 
In the original version of the theory, $\varphi$ is assumed to be spatially constant (corresponding to a homogeneous medium), 
while in the later Landau-Ginzburg form, it is allowed to have a spatial dependence and thus becomes an effective field.
Apart from providing a measure of the amount of order present in a critical system, the significance of order parameters in
modern theory lies in the fact that in many cases they can be thought of as supplying a ``mesoscopic'' level of description, interpolating 
between the microscopic level - pertaining to systems such as atoms and nuclei - and the macroscopic level - pertaining to 
systems such as magnets and fluids. As such, they offer a novel way of viewing complex, condensed matter systems.
All in all, although often underappreciated, Landau's concept of an order parameter brought ``light, clarity and form''
to the general theory of phase transitions and led to an improved understanding of critical phenomena in particular (cf. \citeN{Fisher}).}.
This translates into the restriction of effectively including only those Fourier modes of the order parameter
with wavelengths larger than atomic dimensions.
Expressions for quantities, such as e.g. the free energy, which directly depend on the order parameter, thus also 
involve only these fluctuations. But consistent implementation of this philosophy then requires an integration 
also over all the longer wavelength fluctuations, all the way up to fluctuations of wavelength of the order of
the correlation length (i.e., the characteristic scale below which fluctuations are correlated and which becomes
infinite at the critical point).
Wilson accomplishes this by iteratively splitting up the one large problem - namely, an averaging over fluctuations 
involving many length scales - into a large sequence of small problems - i.e., an averaging over fluctuations involving
only a single infinitesimal length scale - to be tackled one at the time.
This seemingly trivial step is really what is at the basis of Wilson's renormalization group, and, as will now be briefly
illustrated within the specific context of (Euclidean) $\lambda \phi^4_d$-theory, has profound consequences\footnote{The
relevant Lagrangian density here (obviously) is given by the Wick rotated form of (\ref{phi4Lagr}) 
and for $d=3$ corresponds to essentially the form of the free energy proposed by Landau and Ginzburg within the context
of their phenomenological model for superconductivity.
In Wilson's reconsideration of the Landau-Ginzburg free energy, as fluctuations on successive length scales are 
integrated out, new (free) energy functionals, $F_{L + \delta L}$, are generated from old ones, $F_L$.
This thus corresponds to the basic renormalization step and it gives rise to effective differential equations for the
coefficients in the free energy (see also the main text).
For large $L$, it is then usually the case that $F_L$ runs into a fixed point of the transformation and the 
solutions for the differential equations governing the coefficients, evaluated for $L$ equated to the correlation length, 
lead to expressions for the critical exponents. See \citeN{Wilson1} for further discussion and explicit calculations.}.\\
The exclusion of short wavelength \emph{statistical} fluctuations is taken to translate into a similar exclusion of 
short distance (or, equivalently, high energy) \emph{quantum} fluctuations within the general framework of (Euclidean) 
quantum field theory (see however the discussion in subsection \ref{EFTparadigm} for some critical remarks).
In other words, assuming this analogy to be valid, the Wilsonian view strongly suggests that \emph{any} quantum field theory (i.e.,
also if it is renormalizable), comes equipped with an intrinsic short distance cutoff, $\Lambda^{-1}$,
below which the theory should not be trusted.
In generic functional integral expressions such as
\begin{equation}\label{infpathint}
Z(\Phi_1; \Phi_0) \; = \; \int \! \mathscr{D} \Phi \, e^{i S[\Phi]}
\end{equation}
such a cutoff is again implemented as an effective 
restriction on the Fourier modes of the field.
Instead of summing over \emph{all} histories of a generic field $\Phi$, as in (\ref{infpathint}), one thus sums only
over those histories for which the Fourier modes $\Phi(k)$  satisfy $|k| \leq \Lambda$ (symbolically this may be expressed 
by replacing the path integral ``measure'' $\int \! \mathscr{D} \Phi$ in (\ref{infpathint}) by $\int_{\Lambda} \! \mathscr{D} \Phi = \prod_{|k| \leq \Lambda} \int \! d \Phi (k)$).
Wilson's key step now consists of performing any relevant functional integral first over only the momentum slice $b\Lambda \leq |k| \leq \Lambda$,
with $b < 1$ such that this slice is ``sufficiently small'', and to subsequently \emph{absorb the effects of this manoeuvre into
a redefinition of the parameters appearing in the action}.
Thus, concretely for $\lambda \phi^4_d$-theory, carrying out the integration over the selected slice and then rescaling
both the momentum, $k \rightarrow b^{-1}k$, and the field, $\phi$, such that the kinetic term in the action retains its
form, results again in a functional integral of the basic form (\ref{infpathint}) with the same cutoff parameter,
but in terms of an effective density
\begin{equation}\label{Leffphi4th}
\mathscr{L}_{\mbox{\footnotesize eff}} \; = \; - \frac{1}{2}(\partial \phi)^2 \: - \: \frac{1}{2}m'^2 \phi^2 \: - \: \frac{1}{4!} \lambda' \phi^4 \: - \: \xi_1' (\partial \phi)^4 \: - \: \xi_2' \phi^6 \: - \: \cdots
\end{equation}
where the parameters, $m'$, $\lambda'$, $\xi'_1$, $\xi'_2, \cdots$ are well defined functions of the original parameters,
$m$ and $\lambda$, and the momentum integrals in question\footnote{For explicit evaluation of some of these integrals, see e.g. \citeANP{PesSch}, ibid.
The circumstance that a renormalization group transformation - the purpose of which is to eliminate a
length or energy scale from a problem - could produce an effective interaction with \emph{arbitrary many couplings}
without being a disaster, was first established in \citeN{Wilson2}.}.
The point is now that since the higher-order terms associated with the coefficients $\xi'_1$, $\xi'_2$, $\cdots$ (which are
non-renormalizable in $d=4$ spacetime dimensions) are generated already in the first step of integrating over just a single
small momentum shell, it may just as well be assumed that these coefficients (or rather, their unprimed versions) are already present in the initial density.
In other words, the presence of a finite, ultraviolet cutoff, $\Lambda$, suggests that the high-energy behaviour of the 
scalar field is best thought of as being governed by an \emph{effective} Lagrangian density of the form (\ref{Leffphi4th}),
rather than the standard form (\ref{phi4Lagr}).
But this interpretation obviously stands or falls with the possibility of recovering the standard form at relatively low
energies. Assuming perturbation theory to be valid near the cutoff and upon viewing all terms in (\ref{Leffphi4th}) beyond the first 
term as perturbations of that term, it turns out that a single integration results in the following expressions for the
transformed parameters
\begin{equation}\label{coeffRGflow}
m'^2 \; = \; m^2 b^{-2} \qquad \lambda' \; = \; \lambda b^{d-4} \qquad \xi_1' \; = \; \xi_1 b^d \qquad \xi_2' \; = \; \xi_2 b^{2d-6} \qquad \cdots
\end{equation}
where all higher-order coefficients, $\xi_3$, $\xi_4$, $\cdots$, are multiplied with positive, integer powers of $b$ (for $d \geq 4$).
Now, suppose that one wishes to compute some correlation function - i.e., essentially a functional derivative of some
order, of an expression of the basic form (\ref{infpathint}) - for which all external momenta, $p_{\alpha}$, have magnitudes
many orders smaller than $\Lambda$.
In the physically relevant case of $d=4$ spacetime dimensions, integration over a large number of successive 
momentum shells, all the way down to the scale of the external momenta, then causes all non-renormalizable 
interactions in (\ref{Leffphi4th}) to effectively melt away. In this sense, these interactions are thus ``irrelevant'' at low energies.
More precisely, eq. (\ref{coeffRGflow}) implies that $\xi_1 \rightarrow \xi_1 (|p_{\alpha}|/\Lambda)^4$ at the scale
of the external momenta, $p_{\alpha}$, and, similarly, for the higher-order coefficients one has $\xi_i \rightarrow \xi_i (|p_{\alpha}|/\Lambda)^{m_i}$, $i=2,3,\cdots$, $m_i \in \integer{}^{+}$.
Hence, at relatively low energies, the non-renormalizable interactions are severely suppressed - by some positive
power of the ratio of the low- to the high-energy scale, that is.
On the other hand (and actually independent of dimensionality), the mass coefficient grows with each successive
integration and is therefore said to be ``relevant'' (in fact, this feature of the mass coefficient will also prove to be
relevant for the discussion in section \ref{naturalness}).
The ``marginal'' quartic self-interaction is a special case.
It is clear from (\ref{coeffRGflow}) that $d=4$ is a critical dimension and that the quartic interaction term is irrelevant,
resp. relevant, precisely in $d>4$, resp. $d<4$ dimensions.
A higher-order analysis gives that $\lambda$ slowly decreases with decreasing energy in $d=4$ dimensions, to the extent
that it approaches zero upon sending the cutoff to infinity\footnote{After one integration one has in fact 
\begin{equation}\label{WRGlambdaflow}
\lambda' \; = \; \lambda \: - \: \frac{3 \lambda^2}{16 \pi^2} \log 1/b
\end{equation}
See e.g. \citeANP{PesSch}, ibid., sct. 12.1.
Note that the infinitesimal version of (\ref{WRGlambdaflow}), not surprisingly, is precisely the beta function for 
$\lambda \phi^4_4$-theory to lowest nontrivial order (i.e., the left-hand side of (\ref{WRGlambdaflow}), on general
grounds, can be expressed as $\lambda' = \lambda + \mu (b-1) d \lambda / d \mu$ and the claim follows upon expanding $\log b$
on the right-hand side in terms of the deviation of $b$ from unity).}.
In other words, just as what could be established directly on the basis of the foregoing discussion on the beta 
function and eq. (\ref{bareseriesexp1}) (see also subsection \ref{SMRGflow}), a renormalization group analysis in the 
sense of Wilson implies that $\lambda \phi^4_4$-theory is trivial as a continuum theory.
Since this conclusion is obtained under the assumption that perturbation theory is valid in the far ultraviolet regime,
its negation gives that if $\lambda > 0$ is assumed to be sufficiently small in the far infrared regime, 
perturbation theory has to break down at ultrashort distances - in agreement with the findings already obtained earlier by
Gell-Mann and Low.

\subsection{Standard Model Renormalization Group Flow}\label{SMRGflow}

\noindent To lowest order, theories such as $\lambda \phi^4_4$-theory and $\mbox{QED}_4$ have $\beta > 0$
and as seen in the previous subsection, this implies the existence of a Landau pole, i.e., a singularity in the coupling 
strength at finite energy. However, as also remarked, it is possible that higher-order contributions to $\beta$, 
which become increasingly important at short distances and which cause the perturbative approximation to lose its validity, 
have a reducing effect on the growth of the coupling constant, so that the singularity is either shifted to the asymptotic 
realm or, by virtue of the presence of a non-Gaussian fixed point, removed altogether.
Nevertheless, in spite of the lack of control over the non-perturbative region, it is generally believed that no such
fixed point exists for any physically relevant quantum field theory.
For instance, there is some evidence against the existence of a non-Gaussian fixed point in four-dimensional quantum electrodynamics\footnote{Cf. \shortciteN{AdCaGrJa}, \citeN{BaJo}.},
whereas rigorous calculations show that $\lambda \phi^4_4$-theory quantized on a discrete spacetime lattice does
not have a nontrivial continuum limit for zero lattice spacing\footnote{Cf. \shortciteN{FeFrSo}.
It should be remarked here that this statement is valid for $\lambda > 0$, strictly speaking, and that it moreover does
not appear to be conclusive. That is, \emph{if} $\lambda \phi^4_4$-theory nontrivially existed as a continuum theory, a
discrete version of it could be obtained by integrating out short distance degrees of freedom, but this version
would also contain all sorts of irrelevant terms and would not be the theory considered in the calculations mentioned.}.
On the other hand, in the case of a $\mbox{SU}(N)$ Yang-Mills gauge field in four spacetime dimensions, coupled to $N_f$ 
massless fermions in the fundamental representation, the beta-function, to lowest order, equals\footnote{See e.g. \citeN{PesSch}, section 16.7.}
\begin{equation}\label{YMbetafunction}
\beta(g) \; = \; - \frac{g^3}{48 \pi^2} \Bigl( 11 N - 2 N_f \Bigr) 
\end{equation}
and hence is negative as long as $N_f < 11 N/2$.
As there is firm empirical evidence (based e.g. on the decay $\pi^0 \rightarrow 2 \gamma$) that actually $N = 3$ for the 
strong interaction, this means that there could exist up to five additional fermion generations and this interaction would 
still be asymptotically free.
However, unless there exist very heavy neutrinos in which the neutral vector boson cannot decay, the experimental data 
provide strong evidence that $N_f$ cannot be larger than six.
Taking also into account the clear indication offered by Bj{\o}rken scaling, this thus means that, overall, there is 
powerful empirical support that the strong interaction is asymptotically free.\\
As for the weak interaction, it is clear that eq. (\ref{YMbetafunction}) does not completely describe the scaling
behaviour of the gauge coupling in this case, because it fails to take into account the contributions of the Higgs
field.
However, although in general the inclusion of scalar fields also has a destabilizing effect on any ultraviolet Gaussian
fixed point (as may be expected from the short-distance properties of interacting scalar fields considered in isolation),
it turns out that in the specific case of the standard model, i.e., with a $\mbox{SU}(2)$ weak gauge field, $N_f = 6$
and a single complex Higgs field, the beta-function remains negative\footnote{The net effect of the Higgs field is to contribute
a term to eq. (\ref{YMbetafunction}) equal to $1/4$ times the fermion term. See \citeN{GroWil}, section VI.}.
So the $\mbox{SU}(2)$ sector of the standard model is also asymptotically free.
Now, although the likely triviality of $\lambda \phi^4_4$-theory, for instance, can be removed by including non-Abelian
gauge fields in an appropriate way\footnote{See e.g. \citeN{Callaway}.}, this does not appear to work in the particular 
case of the standard model.
In fact, a complete treatment of the scaling behaviour of the Higgs and Yukawa couplings, although straightforward
in principle, is somewhat tedious and will not be presented here\footnote{See e.g. \citeN{tHooft3}.
The beta functions for the nine Yukawa couplings in the absence of the gauge fields are also positive.
In fact, the ``price of asymptotic freedom'' is that for physically realistic conditions (i.e., four dimensions,
positive energy, etc.), it can only be achieved through the inclusion of Yang-Mills type gauge fields.}.
At any rate, it is clear that the $\mbox{U}(1)$ sector of the standard model is \emph{not} asymptotically free and the
upshot of the foregoing remarks is that, in spite of the good ultraviolet properties of the strong and
weak interactions, the standard model as a whole is not asymptotically free and is very presumably also not
asymptotically safe.
This thus means that, most probably and \emph{already purely for reasons of internal theoretical consistency}, 
the standard model cannot be considered a complete theory.

\subsection{The Modern Orthodoxy : Effective Field Theories and Finite Cutoffs}\label{EFTparadigm}

\noindent The verdict that the standard model without a cutoff is very presumably a trivial theory is rather startling.
In fact, although this is seemingly not often realized, if it actually \emph{is} impossible to (nontrivially) define
the standard model as a continuum theory, this in itself would be without precedent in the history of physics.
The point is not that there are no conclusive arguments to go beyond the standard model (there are, of course, if only 
because of the model's absolute silence on the phenomenon of gravity). The issue, rather, is that these arguments in principle have
nothing to do with the question of whether or not the standard model can be consistently defined as a continuum theory.
For instance, long before the final advent of either special relativity or quantum theory, hints causing
physicists to doubt the universal validity of classical mechanics already emerged, but these hints did not pertain to 
any lack of internal theoretical consistency.
And while there \emph{are} internal theoretical grounds to doubt the universal scope of general relativity (or classical field theories more generally), 
the situation here is rather different. Indeed, unlike the probable case for the standard model of particle physics (or in fact any quantum
field theory that is not asymptotically safe), the breakdown of some of general relativity's principles for certain physically relevant solutions,
is not implied for any \emph{finite} spacetime regions, but only at ``points'', i.e., ``at'' the actual spacetime singularities.
By contrast, triviality of a quantum field theory is just another way of saying that that particular theory becomes
inconsistent above some \emph{finite} high-energy cutoff, $\Lambda$.
This means for instance that local (anti)commutativity breaks down at distances below $\Lambda^{-1}$ and while this 
may not seem reason for much practical concern provided $\Lambda$ is large enough, it goes to the very heart of what 
is meant by a quantum field theory (as may be recalled, local commutativity of field observables is for instance crucial 
for prohibiting the use of spacelike correlations between quantum fields for superluminal communication)\footnote{Unfortunately, 
the conceptually odd status of the standard model in this regard is often not well appreciated in the particle physics literature, where it is
typically ``explained'' away on the basis that there are many other instances of effective theories in the history of physics 
and that employment of these is conceptually unproblematic (see e.g. \citeN{Georgi} for an argument to this extent).
For instance, in the limit where all speeds are sufficiently small compared to the speed of light, classical Newtonian mechanics 
provides a consistent, effective description of the relevant physics, which yields experimental outcomes which are in practice
indistinguishable from those obtained through the more fundamental description provided by relativistic mechanics. 
The fact that Newtonian mechanics is an effective description of relativistic mechanics in the limit of sufficiently small
speeds is of course not contested here. The point is rather that the former can fully stand on its own and in particular
does \emph{not} become trivial in the limit where the parameters associated with the more fundamental theory are removed
(sending the speed of light to infinity, one typically recovers standard Newtonian expressions from their relativistic counterparts).
By contrast, although there is no difficulty in viewing the standard model of particle physics as providing a good
effective description of the relevant physics for energies sufficiently small compared to some high-energy cutoff, 
$\Lambda$, associated with a more fundamental theory, the standard model cannot be \emph{defined}, at least at the time 
of writing, without making reference to $\Lambda$; this is just what triviality means.}.\\
In a somewhat related fashion, it is sometimes argued that the likely triviality of the standard model is not really 
such a serious problem, because the model already becomes invalid because of quantum gravity effects anyway, long 
before it encounters any putative Landau pole\footnote{Although the estimate (\ref{Landaughost}) was derived for pure QED,
the received view appears to be that the value of a generic Landau pole would be similarly large (see e.g. \citeN{Weinberg6}; section 18.3).
In fact, in the case of the Higgs field in the minimal standard model, a triviality-type argument can be used to put an 
upper bound on the Higgs mass, $M_H$ (since $M_H^2$ is proportional to the Higgs self-coupling), or, alternatively,
given $M_H$, it is possible to obtain an estimate for the scale, $\Lambda$, above which the \emph{perturbative description}
breaks down. A straightforward, first-order analysis gives
\begin{equation}
\Lambda \; \simeq \; v \, \exp \left( \frac{16 \pi^2}{3} \left( \frac{2v^2}{M_H^2} - 1 \right) \right) \qquad \qquad v \; = \; 246 \, \mbox{GeV}
\end{equation}
so that if the recently found boson at CERN with mass given by (\ref{Xbosonmass}) is indeed the Higgs of the minimal standard model, 
the perturbative description would very conveniently remain valid until energies of order $\sim \; 10^{153} \, \mbox{GeV}$.
It should be noted however that the leading order analysis is somewhat crude (especially in view of the strong sensitivity
of the cutoff on $M_H$) and that a more elaborate, second order analysis already yields a significantly lower value of
$\Lambda$. Furthermore, there seems to be no reason in principle why the inclusion of still more orders in perturbation
theory could not entail a drastic, further lowering of $\Lambda$ (\emph{especially} if the position is taken that the 
evidence for the Landau pole is inconclusive).\label{trivbasedmassest}}.
This is not really a convincing argument however, for two reasons.
First, in the absence of both a theory of quantum gravity and a definite verdict on (the particular value of) the Landau pole, one cannot really be
sure that no such pole is encountered before quantum gravity effects become important.
In particular, since all theories of the Planck scale are presently still at a purely speculative stage, there is no 
guarantee that the Planck (energy) scale is not either significantly increased or decreased as a result of quantum gravity effects, or that 
it remains finite at all.
Even though a significant increase may seem unlikely in view of the fact that a standard renormalization group analysis based on
perturbative quantum gravity implies that Newton's constant increases at short distances (implying a corresponding 
decrease of the Planck energy), things are not really so clear. 
The point is that a nontrivial fixed point in quantum gravity - the existence of which is not infrequently argued for - 
is just another way of saying that Newton's constant converges to a fixed nonzero value at infinitely short
distances and thus raises a paradox. In order to resolve this paradox, some authors have argued that the issue of whether a
fundamental cutoff exists in quantum gravity may depend on the \emph{choice of units}\footnote{Cf. \citeN{PerPer}.\label{perpernote}}.
Quite apart from this there is the fact, that not all present-day standard theories treat the Planck scale as a sacrosanct, 
fundamental cutoff. For instance, the currently fashionable inflationary paradigm explicitly
attributes physical significance to modes \emph{far} beyond the Planck scale (although it is probably fair to say that inflationary
theorists usually regard such a ``trans-Planckian domain'' as a major open problem of their theory).\label{LPlscaling}
A second reason for dismissing the Planck-scale argument against the importance of the Landau pole is simply that gravity 
is external to the standard model and it should therefore be of no relevance, in principle, to the issue of whether 
standard model theories can be defined as continuum theories\footnote{Or conversely, \emph{if} a cutoff of 
Planckian order is actually taken seriously as some sort of \emph{intrinsic} physical cutoff for standard model 
quantum field theories, this should in principle have novel implications for such theories (in the form of 
additional constraints on their physical parameters, for instance).\label{addconstrSMparam}}.
Put somewhat differently, the in itself not unreasonable view that the gravitational interactions ignored by the standard model
become dominant near the Planck scale and that for that reason the model cannot be trusted as a \emph{physical} model
near the Planck scale, has in principle no ramifications for the triviality issue, which is primarily concerned with
the \emph{internal theoretical consistency} of the standard model just by itself (moreover, for the purely asymptotically free components of the
model, the physical argument that gravitational interactions become dominant when reaching Planckian energies
would remain just as valid, but the fact that these components \emph{can} be defined on all distance scales, or so
it is usually asserted, shows that such a physical argument can have no bearing on these components, when considered just by themselves).\\
What can at most be argued is that if the usual value for the Planck scale, i.e., $G_{\mbox{\scriptsize N}}^{-1/2} \simeq 10^{-19} \, \mbox{GeV}$,
and an estimate of the form (\ref{Landaughost}) remain roughly correct (and in view of the previous remarks, these are not trivial assumptions), the Landau pole would pose no 
problem for the standard model \emph{in practice}.
But since this is just a ``lucky coincidence'', which is moreover strongly dependent on the particular values of the physical
parameters in terms of which the Planck scale and the Landau pole are defined, it does not amount to a very satisfactory
explanation (for instance, in the case of pure QED, it is clear that the specific value (\ref{Landaughost}) is highly
sensitive to the actual value of the fine structure constant; if a value for $\bar{\alpha}$ is taken that is only
an order of magnitude larger than its actual value, $\mu_{\infty}$ is already lowered to Planckian orders)\footnote{These observations
could however have significance for another reason, but it would lead to too much of a digression to discuss this here.}.
At any rate, once practical considerations are brought into play, it becomes necessary to contemplate the possibility
that there are \emph{other} physical cutoffs for the standard model, besides the Planck scale.
Although it is true that the Planck scale currently represents the only semi-reliable prediction of the existence 
of a physical scale beyond, say, TeV order, a few elementary facts should be kept in mind.
First and foremost, despite the fact that the present formulation of the standard model could well turn out to be
partially founded upon theoretical artefacts, there appears to be little doubt amongst particle physicists that the model
provides an impressive, approximately correct account of high energy experiments\footnote{As is well known for instance,
the physical meaning of the gauge principle (if any) remains to be clarified.
In this regard, it is interesting to note that according to the formulation of quantum field theory initiated by
Haag and collaborators, coloured quarks and gluons can be extracted from the net of local observable algebras, by 
proceeding to its short-distance scaling limit and without the need to introduce any unobservable gauge degrees of
freedom. 
Within this setting, the quarks and gluons and the global colour invariance are referred 
to as \emph{ultraparticles} and \emph{ultrasymmetries}, respectively, but their physical reality is beyond doubt.
Cf. \citeN{Buchholz}.}.
But this was only achieved as a result of a long, proper dialogue between theory and experiment.
In particular, no one ``ordered'' particles such as the muon or the chiral neutrino, while in other instances, where 
discoveries were anticipated by some researchers, as occurred with e.g. the charmed quark or the tauon, they implied 
a break with prevailing opinions at the time.
To think that there is no new physics beyond the typical standard model scale for at least sixteen orders of magnitude
requires a rather big leap of faith and at any rate is completely unwarranted from a purely experimental point of view.
Closely related to this is the fact that there are of course important reasons other than gravity to think that there 
is physics beyond the standard model of particles.
Many features of that model (e.g., the origins and particular values of physical parameters such as coupling
constants and mixing angles, the specific number and structure of fermion generations, parity violation, etc.)
have an ad hoc quality and some of these only appear to be present to ``save the phenomena''.
Such features seem prima facie unrelated to (quantum) gravity, but clearly cry out for an explanation.\\
Historically, attempts to improve upon the standard model because of these ``internal'' motivations very much arose
in the form of ``top-down'' approaches.
For instance, according to the so-called decoupling theorem, given a renormalizable theory involving two sets of fields, 
$\{ \Phi_i \}$, $\{ \phi_i \}$, governed by widely separated mass scales, $M_i$, $m_i$, respectively, it is in general 
possible to choose a renormalization prescription, such that at low energies, $E \ll M_i$, the contributions of the heavy 
fields $\Phi_i$ are either suppressed by positive powers of $E/M_i$, or are absorbed into renormalizations
associated with the light fields\footnote{Cf. \citeN{AppCar}.}. 
In other words, at sufficiently low energies, the heavy fields effectively ``decouple'' and
physical processes are appropriately described by a low-energy, effective field theory (EFT) for the fields $\phi_i$, with
non-renormalizable interactions suppressed by positive powers of $E/M_i \ll 1$. Intuitively, it would seem that such a
decoupling of heavy fields at low energies in a renormalizable theory is precisely what accounts for the phenomenological 
success of the original Fermi model for weak interactions (i.e., seen as the effective, low-energy $\mbox{SU}(2)$ part of 
the Glashow-Salam-Weinberg model).
Moreover, even though the standard model itself is of course ``renormalizable'', it would also seem that a similar idea
could in principle be employed to attempt to account for the three standard model interactions as different low-energy,
effective manifestations of a single, high-energy interaction associated with a so-called grand unified theory (GUT), 
with a gauge group $G$ that is spontaneously broken into $\mbox{SU}(3) \times \mbox{SU}(2) \times \mbox{U}(1)$ at low energies\footnote{These
remarks are only intended as heuristic and more or less follow the common wisdom on these matters encountered in typical
discussions. In particular, Fermi's theory by itself is just a non-renormalizable theory, the effects of which are not
exactly suppressed at low energies, and which moreover contradicts the frequently expressed viewpoint that all ``low 
energy physics'' is perfectly accounted for by \emph{renormalizable} theories.
Also, electroweak interactions at high energies (before symmetry ``breaking'') are of course not really unified, as there 
are two independent coupling constants, associated with respectively $\mbox{SU}(2)$ weak isospin and
$\mbox{U}(1)$ weak hypercharge.}.
Unfortunately however, spontaneously broken gauge theories do not fall under the scope of the decoupling theorem, but 
even if they did, the fact that predictions like proton decay, based on the simplest possible choices for $G$,
such as \mbox{SU}(5) and \mbox{SO}(10), have so far not been borne out by experiments, suggests that a top-down 
approach in this particular form is probably not feasible (see also below)\footnote{Spontaneously broken gauge 
theories violate some of the assumptions of the decoupling theorem (cf. \citeN{ColWilZee}) and as a result, 
``heavy loops'' do not decouple at low energies in such theories.
In fact, the strong sensitivity of electroweak observables to the value of the top mass (as exemplified for instance by 
the quadratic dependence of loop corrections to the charged vector boson mass on the top mass) is a direct consequence 
of this. Because of non-decoupling, electroweak precision tests could in principle be useful also as probes for physics 
beyond the standard model.\label{decouplingthm}}.\\
Another general class of approaches to move physics beyond the standard model - considerably closer in spirit to
the manner in which the standard model itself actually came to be constructed - can be characterized as ``bottom-up''.
Here, the high-energy GUT is not supposed to be known - or to even exist at all - and the basic idea is to work
progressively towards higher and higher energy scales by extracting more and more information about the
various, a priori possible (renormalizable and non-renormalizable) interactions.
Even though there are in principle infinitely many of such interactions, two simple, physically plausible conditions
guarantee that, at least in perturbation theory, only a finite number of terms is needed to calculate quantities of 
physical interest at energies $E$ smaller than some given mass scale, $M$, to any finite degree of accuracy\footnote{See e.g.
\citeN{Georgi}.}.
As $E$ approaches the scale $M$ from below, the non-renormalizable interactions become ``less non-renormalizable''
as judged from a more encompassing EFT with an associated higher energy scale $M'$ (or strictly renormalizable field theory if $M'$ turns out 
to be infinite), with the old EFT - valid only at energies below or equal to $M$ - and the new EFT related at scale $M$
via so-called matching conditions\footnote{To lowest order, these conditions just express that the appropriate couplings 
in the different EFT's are continuous across the boundary.}.
This way one thus obtains a (possibly infinite) tower of nested EFT's, each theory incorporating more fields than and 
effectively incorporating its lower energy predecessors.
Despite the different underlying motivations of the two approaches towards physics beyond the standard model as just 
sketched, a key feature of both is that \emph{given} a particular EFT with associated mass scale $M$, one does not 
need to know the details of whatever physics is at play at scales larger than $M$, in order to extract physically 
meaningful information from the theory.
Indeed, low energy processes as described by the given EFT are ``effectively'' independent of what goes on at higher
energies, roughly in compliance with decoupling\footnote{With the bottom-up approach, there is a technical caveat in that
the decoupling theorem does not operate in so-called mass-independent renormalization schemes (such as the MS scheme
with dimensional regularization) apparently needed in this approach. Cf. \citeANP{Georgi}, ibid.
However, in this case the EFT is defined such that the ``physics of the decoupling theorem is put in by hand''.}.\\
Concretely, in order to extract useful information about physics beyond the $\mbox{SU}(3) \times \mbox{SU}(2) \times \mbox{U}(1)$
standard model, according to a bottom-up type approach, one should write down all non-renormalizable terms that could
potentially contribute to a given level of accuracy.
The problem however is that $M$ is a priori unknown.
For a given low energy range and level of accuracy, few non-renormalizable operators are needed if $M$ is very large (cf. 
the foregoing remarks) and the fact that non-renormalizable interactions have so far apparently not been required to account 
for experimental results is sometimes used as an argument for large $M$. But this is actually not a very solid argument.
For instance, if only the QED sector of the standard model is considered, one would add the following gauge-, Lorentz-
and CP-invariant, but non-renormalizable operators 
$\bar{\psi} \sigma^{\mu \nu} \psi F_{\mu \nu}$, $(\bar{\psi} \psi)^2$, $(\bar{\psi} \gamma^5 \psi)^2$, $(\bar{\psi} \psi)^3$, $\cdots$.
The first, so-called Pauli term (where $\sigma^{\mu \nu}$ is proportional to a commutator of Dirac matrices $\gamma^{\mu}$, $\gamma^{\nu}$)
 is a dimension-5 operator and is a priori expected to give the leading order correction through its contribution to the 
anomalous magnetic moment, $a^{\mbox{\tiny th}}$, of any charged lepton with mass $m$ an amount of order $8m/M$.
In the case of the electron, there is the well-known phenomenal agreement between theory and experiment expressed by
$|a_e^{\mbox{\tiny th}} - a_e^{\mbox{\tiny exp}}| \lesssim 10^{-12}$, implying that $M \gtrsim 4 \: 10^9 \mbox{GeV}$.
However, if symmetries are present that constrain the form of the non-renormalizable interactions, this lower limit on $M$
significantly reduces (i.e., by a factor of at least $10^4$)\footnote{Cf. \citeN{Weinberg6}, section 12.3.}.
Furthermore, there \emph{is} an almost $4 \sigma$ level discrepancy between measurements of
the muon anomalous magnetic moment and the standard model predictions for it, something which has been translated
into values of $M$ below the Fermi scale\footnote{See e.g. the review on the muon anomalous magnetic moment by
Hoecker and Marciano in \citeN{Beringer}.}.
These lower limits are in sharp contrast with the ramifications of typical GUT scenarios, where a straightforward
renormalization group argument yields for the energy scale, $M$, at which the three standard model couplings 
merge, a value of order $10^{14} \, \mbox{GeV}$ (at least). 
In other words, GUT-type scenarios generically predict the existence of a ``desert'', with no new mass scales and very
little new physics between the electroweak scale and the grand unification scale - i.e. for at least twelve orders of
magnitude\footnote{In some GUT models, the large $G$-symmetry is spontaneously broken in more than two stages, involving
several intermediate high-energy scales, $M_1, M_2, \cdots$ beyond the electroweak scale. As a rule however, the introduction of ``flowers'' in the
desert substantially increases overall complexity and tends to result in a loss of predictive power (cf. \citeN{Langacker}).
In fact, this also applies to supersymmetric (SUSY) versions of GUT's.
Although the introduction of supersymmetry does ameliorate some problems that are typically argued to plage standard GUT's 
(most notably the so-called hierarchy problem; cf. also footnote \ref{GUTnaturalness}) and moreover also does predict (if SUSY is to resolve this hierarchy problem) a new ``superpartner'' mass scale, $M_{\mbox{\scriptsize sp}}$,
well within LHC-range, the minimal SUSY $\mbox{SU}(5)$ GUT model postulates no less than 120 fields and 100 parameters \emph{in addition}
to those already present in the standard model.
As is well known, all GUT scenarios predict that the proton is unstable.
In addition, some simple (non-supersymmetric) GUT's predict a small number of new, low-energy effects 
(such as the existence of a single right-handed, neutral boson with a mass below the TeV-scale in the case of 
the $\mbox{SO}(10)$ model).}.\\
Regardless of which of the two specific implementations of the EFT viewpoint is pursued, it is clear that the viewpoint
itself essentially entails a picture of sub-microscopic physical reality in terms of a hierarchy of effectively
decoupled energy domains and accompanying scales, $\Lambda_1, \Lambda_2, \cdots$, with each domain described by
an effective field theory of some form.
Obviously, the crucial question raised by such a picture is whether the hierarchy is finite and it is with regards
to this question that the two implementations tend to be suggestive of rather different answers\footnote{There is of
course no need to restrict attention to either implementation exclusively (something which indeed appears not to have 
been done in practice). For instance, by starting with a bottom-up type approach, valuable empirical clues could be
obtained with regards to the viability of a generic \emph{class} of top-down approaches (for instance, those that involve some 
particular symmetry at high energies).}.
For present purposes however, this is a bit of an academic issue, since the infinite hierarchy naturally suggested by
(or even implicit to) bottom-up type approaches, seems scientifically empty\footnote{That is, unless the infinitely 
many scales were either a clear prediction or a clear tenet of a consistent theory which could be thoroughly checked
\emph{through other means}, they appear to lack scientific content.
Note however, that a similar remark applies to a sufficiently large finite hierarchy, which should in practice
quickly become experimentally indistinguishable from an infinite hierarchy in a finite-resource world.\label{infhrchyissue}}.
On the other hand, it could well be that the above question is ill-posed and that the hierarchy is (in principle) infinite
in a more abstract, operationalist sense. That is, it could be that no finite underlying theory exists in the traditional
conception of such a theory as providing a consistent, genuinely testable description of physical processes at (and possibly
beyond) some particular high energy scale, which is in some definite sense unique and compelling.
Although it is conjectured here that a finite, more explanatory theory than the standard model does in fact exist, 
the (admittedly slightly rhetorical) question is at the same time raised whether it may not be more probable,
especially in view of the failed efforts of the conventional approaches so far, that the theory in question will fail
to be reductionist in the usual strong sense (see section \ref{disc} for some further remarks). 
Sticking to the typical ``atomist'' mode of thinking common to particle physics however (for the time being), it 
is clear that, modulo the caveat mentioned in footnote \ref{infhrchyissue}, the assumption of a theory more fundamental than the standard model effectively entails a finite hierarchy
and some sort of ultimate \emph{underlying} theory with accompanying, finite high energy scale, $\Lambda_{\mbox{\scriptsize F}}$.
Obviously, retrieving this alleged ultimate theory then becomes a key goal, as well as subsequently relating it to 
empirically accessible phenomena (e.g., via existent or realistically constructible particle accelerators).\\
In its attempts to achieve these goals, high energy physics has unfortunately found itself stuck for the past fourty years or so. 
That is, in spite of the fact that important missing ingredients of the standard model have been successfully
identified in experiments since the construction of the model in the early 1970s (most notably, the weak vector bosons
in 1984, the top quark in 1995, the tau-neutrino in 2000 - and now possibly the Higgs in 2012), \emph{none} of the attempts to go 
beyond the standard model have so far been even remotely successful.
In addition to the (supersymmetric) GUT's already mentioned, a general, even more encompassing top-down approach was 
developped and vigorously pursued for many years in the form of string/M-theory.
In fact, during the 1980s, and especially 1990s, the popularity of this approach was immense and according to its
most outspoken proponents, it was even the ``only game in town''.
However, in view of the googol pentupled of string vacua or so that have been conjectured to exist, string theorists have
now largely abandoned their dream of uniqueness and have moreover become divided over the prospects of string/M-theory
as a physical theory in recent years.
Clearly, a highly nontrivial issue facing any finite high-energy theory in the usual sense, especially if $\Lambda_{\mbox{\scriptsize F}}$
is very large (as happens to be the case in nearly all the conventional approaches), is how to retrieve unambiguous empirical support for it.
This difficulty was already recognized a long time ago, for instance, when prominent particle physicists themselves began
to openly wonder whether the end of their discipline might be in sight (i.e., because of stagnation), but in view of the 
recent LHC findings it has again become highly relevant\footnote{Cf. \citeN{Glashow}.
Although other physicists have taken a more optimistic view regarding the prospects of obtaining experimental support,
it is probably fair to say that a sense of stagnation in high energy physics has long been generally felt (see e.g. \citeN{Weinberg7}
for a plain statement to this effect).}.
In view of these difficulties, two important questions naturally arise. A first question concerns the strength of the 
general case for a finite theory, more fundamental than the standard model in the traditional reductionist sense.
A second, related question concerns the strengths of the individual cases for some of the particular theories
in this regard proposed so far.
Although a proper attempt to answer the second question is, for rather obvious reasons, beyond the scope of the present
treatment, a partial answer for the specific class of supersymmetric theories will be provided in the next section.
More precisely, it will be seen that a particular alleged problem of the standard model, which is widely advertized as requiring 
supersymmetry for its resolution, is not in fact a problem for the standard model \emph{per se}.
With regards to the first question it is clear that if an infinite hierarchy of theories is rejected, the EFT
paradigm \emph{implies} the existence of a more fundamental theory in the sense specified and the first question 
thus translates, essentially, into a question about the strength of the case for the EFT paradigm.
This latter question is addressed in the remainder of this subsection.

\subsubsection*{Difficulties With the EFT Perspective}

\noindent Despite the fact that significant technical and conceptual progress has been made with renormalization 
theory during the past four decades or so - as the discussion in subsection \ref{QFTperspectives} should make clear
- so far the new view on renormalization and quantum field theories that is adequately summarized by 
the EFT paradigm, has not been integrated into a mathematically and conceptually coherent framework.
A number of glaring difficulties and omissions remain.
First, and directly extending the foregoing discussion, the physical significance of the various scales $\Lambda_1, \Lambda_2, \cdots$
remains rather mysterious. \emph{Only} if the first energy scale, $\Lambda_1$, beyond the electroweak scale 
is very large - as typically predicted by any top-down approach that incorporates a (non-supersymmetric) GUT scenario - does one have a 
natural (i.e, Wilson-type) explanation of why the ``low-energy'' realm associated with the standard model is so well 
described by a formally renormalizable theory (as will be seen in section \ref{naturalness}, there is then also a prima 
facie ``naturalness problem'' - i.e.,  an apparent problem of fine-tuning for relevant operators, such as 
scalar mass terms).
On the other hand, if $\Lambda_1$ is relatively small, the renormalizability of the standard model becomes somewhat of
a mystery from a Wilsonian viewpoint.
As will become clear shortly, these remarks in fact relate to a second major (partial) difficulty with the modern view of
renormalization, namely the issue of whether the two different versions of the renormalization group introduced in
subsection \ref{QFTperspectives} are effectively equivalent.\\
The lack of a clear verdict on the status of the renormalizability criterion becomes even more manifest at the opposite
side of the hierarchy. That is, even if it is assumed, in line with the foregoing remarks, that some finite, fundamental 
theory with corresponding scale $\Lambda_{\mbox{\scriptsize F}}$ (which could be equal to $\Lambda_1$) exists ``beneath''
the standard model, the EFT paradigm itself has very little to say on whether $\Lambda_{\mbox{\scriptsize F}}$
is a truly fundamental cutoff in any meaningful sense.
For instance, from a purely field theoretical perspective, $\Lambda_{\mbox{\scriptsize F}}$ could merely represent
an auxiliary, low-energy effective scale associated with some non-renormalizable interaction, obtained by integrating out
high-energy degrees of freedom in some strictly renormalizable continuum field theory - in accordance with the decoupling theorem. 
But, given the changed perspective on non-renormalizable interactions (i.e., their potential full-blown viability; cf.
footnote \ref{supdivnote}), an appeal to decoupling may not be necessary.
Stated more generally, the presence of the various scales within the EFT viewpoint underscores the fact that there are
now two rather different perceptions of non-renormalizable interactions.
On the other hand, if $\Lambda_{\mbox{\scriptsize F}}$ is an absolute cutoff in some definite sense, matters do not appear
to become much clearer.
An absolute cutoff necessarily connects with the properties of space and time and according to our best-informed,
present-day scientific worldview, these properties are described very successfully by general relativity.
Indeed, since according to general relativity, spacetime properties are encoded in the gravitational field by way of the metric,
the idea that concepts such as length and area simply loose physical meaning at the cutoff acquires a more profound significance.
Yet, taking seriously the particle physics argument that gravity becomes dominant near the scale $\Lambda_{\mbox{\scriptsize F}}$ 
(for which the only natural candidate obviously is the Planck scale, $G_{\mbox{\scriptsize N}}^{-1/2} \simeq 10^{-19} \, \mbox{GeV}$),
then implies that the curvature of spacetime becomes very large near the cutoff.
In fact, it seems highly likely that spacetime curvature already becomes non-negligible several orders of magnitude
below the cutoff, so that, quite apart from the issue of what form the fundamental theory describing physics at the
cutoff takes, it becomes necessary to appeal to the theory of quantum fields in curved spacetime
(in the absence of a theory of quantum gravity that should allow the inclusion also of back-reaction effects).
As is well known however, many of the special features of flat spacetime quantum field theories - in terms of which
the EFT paradigm is exclusively formulated - become lost or altered upon replacing the flat background spacetime by
a general curved background (cf. footnote \ref{QFTsubtleties}).\\
A second major difficulty with the new view on renormalization originates in several related, unresolved ``issues of
equivalence''. For instance, although it is clear that the two distinct renormalization group approaches pioneered by 
Gell-Mann and Low and by Wilson, respectively, share a number of crucial characteristics, opinions diverge as to whether 
these two approaches are in fact equivalent in any meaningful sense of the word.
What seems to clearly point to \emph{in}equivalence is that, in a Wilson-type approach, with its emphasis on distinct
physical energy scales and explicit short-distance cutoffs, there is, as just mentioned, an a priori problem of fine-tuning
for relevant operators, whereas it will be seen in subsection \ref{custosymm}, that in a renormalization group approach based on dimensional regularization 
(supplemented by e.g. the MS prescription), there is no such problem (and indeed, in such schemes no \emph{explicit} 
cutoff is ever introduced)\footnote{On the other hand, in so-called ``mass independent renormalization schemes'' 
(such as MS in dimensional regularization) the decoupling theorem does not hold, so that it seemingly cannot a priori be argued 
- in contrast to a Wilsonian approach - that all irrelevant operators are highly suppressed at low energies.
Because of the inapplicability of the decoupling theorem in such situations, some authors have contrasted EFT's 
based on Wilson's renormalization group with ``continuum EFT's'' and argue that these are inequivalent (\citeN{Georgi}).
For discussions and different views on the issue of equivalence of the Gell-Mann and Low and Wilson approaches, see e.g.
section 4 of \citeN{Cao} and references therein (in particular footnote 14 and the concluding remarks on p. 16), \citeN{Wilson2},
\citeN{Weinberg2}.}.
A related issue is that Wilson's approach relies more on classical intuitions - inspired as it was by the wish to
better understand the puzzling behaviour of critical systems. It appears that arguments for the equivalence
of the two renormalization group approaches (tacitly) assume that quantum field theory is essentially nothing but
a sophisticated version of statistical mechanics.
This thus raises a related, but distinct issue of equivalence, namely whether quantum field theory is equivalent - in a
manner yet to be clarified - to (a sophisticated version of) statistical mechanics.
In fact, there are two separate issues here, since in arguments of this variety, quantum field theory is actually 
understood in an Euclidean sense (i.e., as defined on a flat Euclidean four-dimensional manifold).
However, although path integral expressions for Schwinger functions (i.e. Euclidean Green's functions) can indeed be rotated 
back to Feynman amplitudes satisfying the usual Wightman axioms if certain reasonable conditions are met, 
opinions appear to diverge on whether these (so-called Osterwalder-Schrader) conditions are actually valid\footnote{See \citeN{Haag}, sct. VIII.1, for a discussion.}.
Furthermore, and of direct relevance to the above remarks on the need to generalize quantum field theory
in the case of an absolute cutoff, the notion of a Wick rotation breaks down in a general curved spacetime\footnote{More
precisely, Wick rotations are not defined for non-static spacetimes. Cf. \shortciteN{AmLoRo}.}.
Regarding the purported equivalence of Euclidean quantum field theory and statistical mechanics, a crucial distinction
between the two - as typically employed in practical cases - is the following.
In the case of a statistical system, a scale-dependent physical variable, such as the free energy, $F_L$, typically
runs into a fixed point for large $L$ (i.e., in the low energy regime, in standard particle physics jargon) and this
moreover corresponds to critical behaviour - as reflected in the presence of anomalous dimensions\footnote{For a detailed
account, see \citeN{Wilson1}.}.
In high-energy applications of quantum field theory on the other hand, no infrared fixed point exists (or is even
desired), while in the deep ultraviolet regime, if a fixed point exists at all, it is far from clear that this has 
anything to do with criticality (apart from the formal analogy based on asymptotic scale invariance as a result of 
the anomalous dimensions of field operators). This is related to the fact that in quantum field theory, the 
renormalization group is taken as a means to assess the influence of quantum fluctuations at \emph{ultrashort distances}
- and thus differs considerably from the way it was originally propounded by Wilson as a method
to deal with fluctuations at all scales, from atomic distances right up to the correlation length\footnote{Cf. \citeN{Wilson1}.}.\\
Within the framework of axiomatic - or constructive - quantum field theory, the view of Euclidean quantum fields as
(generalized) random variables can be rigorously justified if the so-called Nelson-Symanzik positivity condition
pertains. Yet, although for some models - such as that of the free scalar field - this can indeed be shown to be
the case, it does not appear to be known whether the condition holds for general models\footnote{Cf. \shortciteN{FeFrSo}.
In the original approach pursued by Symanzik and Nelson, the said positivity condition is a consequence of the explicit 
definition of quantum fields as random variables on a probability space.}.
Finally, in statistical mechanics, the whole renormalization group approach is based on the recognition that a definite
physical cutoff - the atomic scale - exists, which is in some sense ``fundamental'' and which is moreover firmly established
on empirical grounds.
Again, although arguments for a fundamental cutoff to spacetime in the form of the Planck scale certainly exist, 
as seen earlier, these arguments are not conclusive (cf. footnote \ref{perpernote}) - let alone supported by any experimental evidence.
Thus, from this angle, the equivalence issue in the end also leads back to the first main difficulty with the EFT paradigm 
encountered above, viz. the issue of the physical meaning of the scales $\Lambda_1, \Lambda_2, \cdots$ within this paradigm.
However, even if the existence of a fundamental spacetime cutoff is assumed - so that one is essentially forced
to deal with a lattice field theory - any purported equivalence between quantum field theory and statistical mechanics
certainly does not become manifest.
For instance, it is well known that a lattice formulation of quantum gauge field theories precludes
the incorporation of basic physical phenomena such as spontaneous symmetry breaking \footnote{According to \citeANP{Elitzur}'s
\citeyear{Elitzur} theorem, all non-vanishing correlation functions must be gauge invariant. See also \shortciteN{DDG}.
Hence, a standard scalar field cannot pick up a nonzero vacuum expectation value.
More generally, within the manifestly gauge-invariant formalism of lattice gauge theories, all correlation functions involving charged fields necessarily vanish. 
Cf. \citeN{Strocchi}, section 4.3. Incidentally, the inference should not be drawn from the foregoing remarks that the ``atomic lattice''-based renormalization 
group of statistical mechanics is free of problems. For instance, also here there is no guarantee that the renormalization group
transformations posess fixed points or are even mathematically well defined (in fact, it appears that obstructions
to a mathematically rigorous theory are related to difficulties with so-called Gibbs measures).
See e.g. \citeN{Wilson1} and references therein.}.\\
Apart from the physical meaning of the various high-energy scales, a deep conceptual difficulty with the
EFT paradigm - or, in fact, renormalization theory more generally - concerns the meaning of the scale
\emph{dependence} of physical observables (as is typically argued for).
In fact, the dependence of physical parameters, such as masses and coupling constants - or more precisely,
the values of physical observables after renormalization - on the scale ``at which they are probed'', is conceptually
problematic for a number of reasons - most importantly, because in the absence of any definite statement about the limiting
behaviour of such scaling, ``elementary particles'' are essentially deprived of intrinsic properties such as mass, 
charge and so on, which they are a priori expected to have in a typical reductionist account of physics.\\
Usually, the scale dependence is argued to be physically real because it (a) in fact is observed in experiments 
and (b) is physically explained in terms of (anti-)screening effects that occur as a consequence of vacuum polarization.
Now, regarding the first argument, it is \emph{not} contested here that in order to successfully account for high-precision
particle scattering experiments, it is necessary to use different values for renormalized coupling constants, depending
on the energy scales at which such experiments operate and in seeming precise accord with the predictions of renormalization
theory.
For instance, taking the classic example of QED, eq. (\ref{QEDRG}) predicts that, to lowest order,
$\alpha^{-1} (\mu \simeq 100 \, \mbox{GeV}) = \bar{\alpha}(1 - 10 \bar{\alpha}/3 \pi \, \ln 10) \simeq 134.6$,
after equating the reference scale $\bar{\mu}$ to the electron mass $\sim 0.5 \, \mbox{MeV}$ (in reality, it is necessary to 
take into account the renormalization effects of all possible charged particle types and when this is done, the effective 
fine structure constant becomes somewhat larger still; $\alpha^{-1}_{\mbox{\scriptsize eff}} (100 \, \mbox{GeV}) \simeq 128.9$,
in very good agreement with precision data from electron-positron colliders, like LEP and others\footnote{The quoted value
for $\alpha^{-1}_{\mbox{\scriptsize eff}}$ corresponds to what appears to have been the standard value until fairly
recently (see e.g. \citeN{Weinberg6} or \citeN{alephcs}) : a slightly smaller value of 127.9 is reported by the Particle Data Group (see the review by Erler \& Langacker in \citeN{Beringer}).
The reason for setting $\bar{\mu}$ equal to the electron mass is that this corresponds to the scale where renormalization 
effects start to occur. That is, the typical range of radiative corrections to the Coulomb potential is the 
Compton wavelength of the electron (i.e., the lightest particle that couples to the photon). 
See e.g. \citeN{PesSch}, section 7.5 (note that this interpretation of 
the Compton wavelength fits naturally with the more general view of it as the typical range in which field fluctuations
are still correlated).
Thus, for distances larger than $\sim 10^{-12} \, \mbox{m}$, the fine structure constant has its familiar value, 
$\bar{\alpha} \simeq 1/137$, while for distances smaller than this, it effectively increases with decreasing distance,
as described to lowest order by eq. (\ref{QEDRG}).}).\\
What \emph{is} called into question here however is the standard explanation of this dependence of effective couplings
on scattering energies in terms of (anti-)screening effects due to clouds of virtual particles.
As already pointed out before, within the context of quantum field theory, the notion of a (real) physical particle 
strictly speaking only makes sense in the absence of interactions (and in stationary spacetimes).
In the usual perturbative approach to quantum fields, this difficulty is sidestepped by the assumption that interactions
merely generate minute corrections to essentially free particle behaviour.
But, apart from the elementary fact that such an assumption is not exactly corroborated by the quantum field theory
formalism \emph{an sich} (i.e., due to the need for renormalization, the expected divergence of the renormalized
perturbation series and the Landau pole problem - features all relevant for the standard model), there is the point 
already made before, that particles in quantum field theory cannot in general be conceptualized as being suitably localized.
Yet, such conceptualization is paramount to any ``physical explanation'' of scale dependence of coupling constants
in terms of charge screening. In particular, states that can be characterized by typical particle properties,
such as sharp energy and momentum, are highly nonlocal and while it \emph{is} possible to construct strictly localized
states, the localization in question is relative to a particular Lorentz frame\footnote{The issue of localization
in quantum field theory has a long and complicated history and will not be discussed here.
A basic problem is that there is no position operator in quantum field theory, thus precluding the usual definition
of a wavefunction via the position representation.
Apart from the mentioned ambiguity mentioned in the main text, the localized Newton-Wigner wavefunctions also suffer
from (superluminal) spreading. See e.g. \citeN{Ruijsenaars}.}.
Moreover, although for a particle of mass $m$, the ambiguity in defining the localization is of the order of the
particle's Compton wavelength, $m^{-1}$, - and thus in a sense is only small for a massive particle - as just noted,
this is exactly the scale at which the coupling constants start to ``run''.
In other words, the irony is that although massive relativistic particles can be approximately localized even in quantum
field theory, this notion of localization breaks down at distance scales smaller than the Compton wavelength -
precisely where it is needed to conceptually account for charge (anti-)screening.
Indeed, the pictorial explanation usually given of why a ``test probe'' effectively sees a larger charge when approaching,
say, an electron at small distances, is that of a sharply localized electron surrounded by a ``cloud'' of sharply localized,
virtual electron-positron pairs. The latter are envisaged to screen the electron's ``true charge'', $e_0$, and this effectively
produces ``the'' electron charge, $e$ (satisfying $4 \pi / e^2 \simeq 137$), which is thus appropriate only at distance scales larger than 
$m^{-1}_e \simeq 10^{-12} \, \mbox{m}$, while at distances smaller than this, the charge ``seen'' by a test probe
increases more and more, as the probe penetrates further and further into the virtual $e^+e^-$-cloud (for accounts
of this variety, see virtually any textbook on quantum field theory or particle physics).\\
The upshot of these remarks is that the usual intuitive picture of running couplings in terms of charge screening
is not only not warranted by the quantum field theoretical formalism; it is specifically spoken against by it.
This is even more so since everything at present indicates that $e_0$ is not a finite quantity.
Thus, even if the localization issue were ignored and the intuitive picture were happily pursued, current evidence
points against the ``true charge'' of the electron even being a meaningful quantity (to argue that this is of
no importance in view of the fact that $e_0$ can never be ``measured'' in any real experiment, would of course be 
completely besides the point~: it is hard to see how meaning could be attached to intrinsic properties of particles
if they are of infinite magnitude). 
Furthermore, even though particle masses can to some extent be viewed as coupling constants (associated with
self-interaction two-vertices), they cannot be viewed as ``charges'' - there is no such thing as ``negative
mass'' - and an account of the apparent scale dependence of mass in terms of screening by particle-antiparticle pairs 
is thus not available.
In the light of these observations - again not disputing any experimental findings - a crucial unresolved issue thus 
concerns the meaning of scale dependence, in particular, in what sense it can exactly be said to be ``physically real''.

\section{The So-Called Higgs Naturalness Problem}\label{naturalness}

\noindent Even though the case for a finite theory with some corresponding high-energy scale, $\Lambda_{\mbox{\scriptsize F}}$, 
``beneath'' the standard model is very far from conclusive (in particular because of the difficulties associated with the 
EFT paradigm that were just signalled), a number of features of the standard model - most notably
its expected triviality and its neglect of gravity - still entail the need to search for a theory that is in some definite sense ``deeper''.
An obvious question which then presents itself is whether the standard model \emph{itself} has any clues to offer
with regards to the particular form of such a theory.
Although different affirmative answers to this question have been proposed over the years, there appears to be a widespread 
belief, at least until recently, that the alleged difficulty of naturalness with the Higgs mass already mentioned several 
times, is most naturally resolved through the introduction of supersymmetry.
As will be seen in this section however, the difficulty in question is not in fact a problem intrinsic to the standard model :
it only arises upon making some \emph{very specific} assumptions on how to extend physics beyond that model.
Furthermore, when taken to its logical conclusion, one of these assumptions essentially calls into question the 
predictive content of the entire standard model as presently understood and is therefore not viable as such.
In order to avoid unnecessary technical complications, these claims are now demonstrated for $\lambda \phi^4$-theory,
which as noted in subsection \ref{phi4renorm}, essentially represents the pure Higgs sector of the standard model.

\subsection{Loop Corrections to Scalar Masses}

\noindent In terms of Wilson's renormalization group discussed earlier, the mass term in the Lagrangian density for the Higgs field is a relevant operator.
Indeed, as is evident from eq. (\ref{coeffRGflow}), the mass coefficient in the effective density (\ref{Leffphi4th}) that is
supposed to describe physics at the cutoff, picks up a factor $b^{-1} > 1$ with each successive momentum shell of thickness $b$ that is 
eliminated and thus becomes increasingly important upon integrating out more and more short-distance degrees of freedom.
In terms of the mathematically more transparent treatment of renormalization presented in subsection \ref{phi4renorm}, this
feature of the mass coefficient directly reflects the fact that the scalar self-energy, $\Sigma$ (cf. eqs. (\ref{phi4-2point2})-(\ref{renormcond})),
is quadratically divergent within an explicit cutoff scheme - as indeed follows from eq. (\ref{superficial}).
That is, based on a quick power counting argument, the bare mass, $m_0$, receives loop ``corrections'', $\delta m$, 
that depend quadratically on an explicit cutoff, $\Lambda$, so that the physical mass, $m$, satisfies an equation of
the form (cf. eq. (\ref{renormcond}))
\begin{equation}\label{scalarloopcorr}
m^2 \; = \; m^2_0 \: + \: \delta  m^2 \; = \; m^2_0 \: + \: \xi (m / \Lambda) \, \Lambda^2
\end{equation}
with $\xi$ some finite, dimensionless function of approximate order unity.
The usual argument is now that in the specific case of the Higgs particle, there is an enormous fine-tuning problem
if $\Lambda$ is very large. Indeed, taking a value for $m$ just below TeV order, eq. (\ref{scalarloopcorr}) implies that
the (squared) bare Higgs mass is apparently fine-tuned to some 32 decimal places if the cutoff is at the Planck scale.
The fact that such ``conspiratorial'' behaviour seems physically highly implausible, is referred to as the
\emph{Higgs naturalness problem} or \emph{(gauge) hierarchy problem}\footnote{Arguments to this effect can be
found in many references (see e.g. \citeN{Altarelli} or a typical modern textbook on quantum field theory or string theory).
Quite generally and regardless of whether the newly found boson with mass (\ref{Xbosonmass}) is actually the Higgs, it 
is generally accepted that if the standard model picture of electroweak symmetry breaking is correct, the Higgs  mass 
cannot be larger than TeV order because of general unitarity constraints (recall also footnote \ref{trivbasedmassest}).
It should be noted that while (\ref{scalarloopcorr}) was derived on the basis of naive power counting (which is not
necessarily representative of actual short-distance behaviour; recall footnote \ref{supdivnote}), an explicit first order 
calculation indeed gives
\begin{equation}\label{scalarmassrenrm}
\delta m^2 \; = \; i \Sigma_{\Lambda} \; = \; \frac{\lambda}{16 \pi^2} \int^{\Lambda}_{0} \! dk \frac{k^3}{k^2 + m^2} \; \simeq \; \frac{\lambda}{32 \pi^2} \Lambda^2
\end{equation}
where the integration runs over invariant Euclidean momenta.
In fact, the self-energy of any boson or fermion field in four dimensions superficially diverges quadratically, respectively linearly,
but in the presence of gauge symmetry, the actual degrees of divergence of Yang-Mills and fermion field self-energies 
are both lifted to zero (corrsponding to logarithmic divergence), whereas that of a scalar field is unchanged; in the 
latter case, the gauge symmetry structure is thus apparently weaker, in the sense of being unable to further constrain 
the divergent scalar self-energy.
It is also worth stressing that even though $\lambda \phi^4$-theory represents essentially just the part of the
standard model exclusively involving the Higgs field and thus fails to take into account all self-energy contributions
due to couplings of the Higgs to other fields, an equation of the form (\ref{scalarloopcorr}) is valid as a general
expression for the Higgs self-energy. For instance, a full standard model analysis gives that the leading order
contribution to $\delta m^2$ comes from virtual top loops and can be expressed according to
\begin{equation}
\delta m^2 \; \simeq \; - \frac{3}{4 \pi^2} \left( \frac{m_t}{v} \right)^2 \Lambda^2
\end{equation}
which thus corresponds to eq. (\ref{scalarloopcorr}) for $\xi \sim 10^{-2}$ (see e.g. \citeANP{Altarelli}, ibid.).}.
Although the argument just presented seems fairly straightforward, some closer inspection of it - especially against
the background of the discussion in the previous two sections - immediately reveals some serious shortcomings.
First, there is no evident, serious problem of naturalness \emph{internal} to the standard model. 
Indeed, if the model is truly renormalizable (no Landau pole), it has no high-energy scale intrinsically attached to it 
and there is no reason not to remove the cutoff, $\Lambda$ - which after all represents just an arbitrary, intermediate
regulator, to be disposed of at the final step of the renormalization process (cf. step (iv-c) in subsection \ref{renormalization}).
According to a completely analogous argument, the ``natural'' value of the Higgs mass would then be infinitely great.
But the ``problem'' of naturalness is then essentially nothing but a restatement of the original infinities problem of quantum 
field theory, i.e., the problem that, at face value, theoretical predictions for physical observables lead to completely 
wrong answers (namely, infinity) implying that either the theory in question is fundamentally wrong or incomplete, or that 
the predictions are wrongly extracted.
Now, as should be clear from the earlier discussion, originally, for theories such as QED, the standard
view was that such theories were essentially correct descriptions of the relevant physical processes (at least for the
energies then under consideration), but that the naive, straightforward interpretation of calculations made within the
theory, should be modified.
Indeed, this view culminated in the theory of renormalization, which in the particular case of QED, has led to 
phenomenally accurate predictions.
Regarding the value of the Higgs mass however, it is typically argued that the wrong (but ``natural'') prediction in this
case implies that the \emph{theory} is fundamentally incomplete.
Although the incompleteness of the standard model is of course not questioned here (cf. subsections \ref{SMRGflow}, \ref{EFTparadigm}),
the point here is simply that, within the present context of assumed renormalizability, it is hard to see the logic
of many typical discussions, in which (infinite) renormalization is first advocated as being essentially unproblematic 
and justified on the basis of the phenomenal accuracy of theories such as QED, and in which it is subsequently argued 
that the quadratically divergent nature of the Higgs self-energy poses a serious conceptual problem that can only
be resolved if new physics is hiding around the corner, waiting to be discovered by the next round of more powerful
accelerators.
On the other hand, if, as expected, the continuum standard model is indeed trivial and if the estimate (\ref{Landaughost}) 
for the cutoff scale is roughly accurate, the model does have a prima facie, internal naturalness problem, but that problem would clearly 
be overshadowed by the internal consistency problem associated with the Landau pole\footnote{Note that if the exponent 
in (\ref{Landaughost}) were significantly increased by higher order corrections, say, to beyond $\sim 10^3$, there would 
also be a serious internal naturalness problem for fermion masses (although of course still much less so than for
scalar masses).}.\\
As already observed however, it is widely held that a Landau pole would pose no serious threat to theoretical consistency
\emph{in practice}, because of reasons \emph{external} to the standard model.
In particular, it is commonly believed that there exists some ``physical cutoff'' to the standard model, at which new 
particles and interactions make their appearance.
Even though it was also noted that (because of the lack of any definite statements regarding the precise values of both 
the Landau pole and the scale at which the new physics is supposed to enter) this view of things corresponds more to a 
possible working hypothesis than to anything else and is moreover quite irrelevant for the issue of whether 
or not the standard model can be consistently \emph{formulated} as a (nontrivial) closed model, it is this view that 
(tacitly) underlies typical naturalness arguments.
Suppose then that the standard model remains valid up to some scale, $\Lambda_1$, in the sense that for energies larger
than $\Lambda_1$, it is no longer representative of the actual physics that goes on at these energies.
So, the cutoff, $\Lambda$, in eq. (\ref{scalarloopcorr}) represents not just an intermediate regulator, but an actual physical 
scale, i.e., $\Lambda_1$, and according to the usual line of reasoning, this scale cannot be too large if a naturalness problem
is to be avoided.
But since extremely little is known about the theory that describes the putative new physics, $\Lambda_1$ could just be 
an effective scale obtained by integrating out short-distance degrees of freedom in some fundamental, truly renormalizable
theory (in compliance, essentially, with decoupling).
In that case, all calculations of physical observables from first principles - including that of the Higgs mass - should
proceed within the context of the high-energy renormalizable theory, for which, as was just seen, a problem of naturalness does not exist.
Of course, there is no a priori reason why $\Lambda_1$ should have such a status.
There could be a whole hierarchy of short-distance cutoffs, $\Lambda_1, \Lambda_2, \cdots$, pertaining to theories beyond 
the standard model, in accordance with the generic EFT paradigm discussed earlier.
It was also noted however, that within the usual philosophical context relevant to particle physics, the assumption that 
a theory more fundamental than the standard model exists basically implies a finite hierarchy and some sort of ultimate 
\emph{underlying} theory with accompanying, finite high energy scale, $\Lambda_{\mbox{\scriptsize F}}$.
The issue of whether there is a prima facie naturalness problem then essentially boils down to the issue of the
particular form of such a theory.
If again, this form is truly renormalizable and $\Lambda_{\mbox{\scriptsize F}}$ represents just an auxiliary, effective
scale obtained by integrating out heavy fields, there is again no naturalness problem associated with (\ref{scalarloopcorr}).
In particular, although $\Lambda_1$ would still represent an effective physical cutoff for the standard model, the wrong
``natural'' value of the Higgs mass that would arise upon taking $\Lambda_1$ sufficiently large, would just be an 
artefact of the calculations made within the particular effective theory, i.e., the standard model.
This thus demonstrates that in the usual arguments, \emph{it is (tacitly) assumed that the underlying,
fundamental theory is not a continuum field theory}\footnote{It should be stressed that this conclusion strictly
pertains to the usual naturalness arguments, which are based on an equation of the form (\ref{scalarloopcorr}).
For instance, it is generally accepted that (continuum) GUT-type models suffer from a naturalness problem and although
this problem is sometimes identified with the alleged Higgs naturalness problem discussed here, the two problems are actually different.
The basic issue in the GUT context is that some large, grand unified gauge symmetry is supposed to be spontaneously broken
in at least two stages, namely near the GUT scale, of order $10^{14} \, \mbox{GeV}$ (at least), and near the electroweak scale,
$\sim 10^2 \, \mbox{GeV}$. It is the wide separation of these scales that is argued to place highly unnatural constraints 
(i.e., in the form of severe fine-tunings) on the parameters appearing in the potential term of the complete GUT action (although,
at a deeper level, it does not seem entirely clear why this should be, unless these parameters are attributed a clear
physical significance). See e.g. \citeN{ORaifeartaigh}.
This naturalness problem - the magnitude of which is quadratic in the difference of the scales and hence at least of
order $10^{24}$ - is thus envisaged to arise already at the classical, i.e., ``tree'' level, and is therefore
quite different from the (usual) naturalness problem based on an equation of the form (\ref{scalarloopcorr}), 
which crucially depends on the inclusion of loop corrections to classical parameters.
In fact, a naturalness problem of this latter type is argued to occur in GUT-type models as well, in the sense that the fine-tuning
required at tree level is in general destroyed by the inclusion of radiative corrections (cf. \citeANP{ORaifeartaigh}, ibid.).
Upon embedding the standard model within a GUT context, the alleged problem of naturalness based on (\ref{scalarloopcorr})
becomes identical to this second naturalness problem, but presenting the argument in this form obviously results 
in a considerable loss of generality (in fact, the situation in this regard bears some resemblance to the generic prediction
of magnetic monopoles by GUT models and subsequent arguments that the lack of experimental evidence for such monopoles
constitutes a serious problem for physics, requiring something like inflationary cosmology for its resolution : quite obviously, the 
absence of empirical evidence for monopoles of GUT-type merely represents an \emph{internal} problem for GUT models).
Incidentally, global supersymmetry is typically invoked as a solution to the (putative) naturalness problem arising
from radiative corrections, but within the context of a GUT scenario, it does not provide an explanation of the required 
fine-tuning at tree level (although it appears that some progress has been made in this direction by \emph{gauging} 
the supersymmetry - something which immediately establishes a link to string/M-theory, since ``supergravity'', i.e., a 
field theory with local supersymmetries, in eleven spacetime dimensions is conjectured to be retrievable as a particular
limit of the hypothetical M-theory).
Another issue is that generic GUT models suffer from a second prima facie naturalness problem, namely the separation
of some twelve orders of magnitudes between the masses of the Higgs colour triplet and the Higgs weak doublet that together 
form the multiplet that picks up a nonzero vacuum expectation in the final stage of the symmetry breaking process, i.e., the transition
$\mbox{SU}(3)\times\mbox{SU}(2)\times\mbox{U}(1)_{\mbox{\tiny Y}} \rightarrow \mbox{SU}(3)\times\mbox{U}(1)_{\mbox{\tiny EM}}$.
This so-called doublet-triplet splitting problem is also widely believed to require supersymmetry for its resolution.
As already noted however, SUSY GUT's in general introduce many additional fields and parameters.
In combination with the property that, in view of the recent LHC findings, many of these new parameters require some
serious fine-tuning themselves, this makes it rather doubtful that supersymmetric grand unification can provide a solution
to any (alleged) fine-tuning problem of the standard model.\label{GUTnaturalness}}. The fact that such an assumption - even though possibly correct - 
is indispensible, obviously weakens these arguments considerably.
Similar remarks in fact apply to other types of arguments that lead to an essentially identical conclusion (cf. footnote \ref{GUTnaturalness}).
In addition, it is also not difficult to see that there is a prima facie Higgs naturalness problem \emph{only} if it assumed
that the bare parameter, $m_0$, in eq. (\ref{scalarloopcorr}) actually carries physical meaning, which is in principle
determined by (or even explicitly calculable from) the fundamental underlying theory with associated scale $\Lambda_{\mbox{\scriptsize F}}$.
Under these conditions, the a priori natural value of $m_0$ would be of order $\Lambda_{\mbox{\scriptsize F}}$,
and the fact that eq. (\ref{scalarloopcorr}) implies a cancellation of any terms of order $\Lambda^2_{\mbox{\scriptsize F}}$
between $m_0^2$ and $\delta m^2$ in order to end up with a physically acceptable value (i.e., of order $10^2 \, \mbox{GeV}$) 
for the Higgs mass, then indeed seems rather contrived if $\Lambda_{\mbox{\scriptsize F}}$ is sufficiently large.
It should be noted however that it is not so clear, especially in the absence of the underlying theory, in what
non-vacuous sense the meaning of $m_0$ could be ``physical''.
In fact, upon recalling the discussion on renormalization in section \ref{QFTinfinities}, it is
difficult to avoid the conclusion that the designation of the required cancellations as problematic essentially amounts
to a rejection of renormalization, which is just expressed in different words (indeed, when viewed from
this perspective, renormalization without a cutoff would have a far more serious ``problem'' of naturalness,
as the cancellations in this case occur to infinitely many decimal places).\\
Of course, as the reader will recall from section \ref{SMtriv} in particular, the position on renormalization in
the present article is fairly critical, while it can moreover not be excluded that bare parameters do actually obtain a 
non-vacuous physical significance within a deeper underlying theory.
In addition, there does appear to be an important difference in the \emph{size} of radiative corrections to the Higgs mass
as compared to analogous corrections to other observables, because of the different dependences of radiative corrections on 
an explicit cutoff (i.e., quadratic versus logarithmic, respectively).
Taking these considerations into account then leads to the impression that, \emph{under the conditions stated} and for
a cutoff of GUT or Planckian order, there is a (unique) Higgs naturalness problem after all.
As will become clear next however, this impression is simply false.

\subsection{Scale Invariance as a Custodial Symmetry}\label{custosymm}

\noindent Quite generally, a very small value of a particular physical parameter, $\alpha$, is said to be (technically) \emph{natural}, 
if setting $\alpha = 0$ has the effect of enhancing the symmetry of the theory, such that the additional symmetry ``protects'' radiative corrections from 
generating a nonzero value for $\alpha$. The symmetry in question is then usually referred to as a \emph{custodial symmetry}\footnote{See e.g. \citeN{Weinberg1}, \citeN{tHooft2}.}.
For instance, the fact that the electron mass, $m_e$, is very tiny relative to the Planck mass, i.e. of order $10^{-22}$,
can be understood in these terms, since setting $m_e =0$ in the QED action leads to an additional $\mbox{U}(1)_{\mbox{\scriptsize L}} \times \mbox{U}(1)_{\mbox{\scriptsize R}}$
chiral symmetry, which forbids the generation of nonzero $m_e$ upon including perturbative quantum corrections.
This is usually explained in terms of the fact that electron mass renormalization is ``multiplicative'',
i.e. radiative corrections to the electron mass in QED are proportional to the bare mass and thus cannot generate a
nonzero mass if the bare mass vanishes.
On the other hand, mass renormalization in $\lambda \phi^4$-theory is usually claimed to be ``additive'' (cf. eq. (\ref{scalarloopcorr})
in the case of an explicit cutoff scheme), in the sense that radiative corrections are not proportional to (some
power of) the bare mass, but instead add to it, effectively. In other words, according to the usual treatments, there can
be no custodial symmetry in this case and one is thus confronted with a problem of naturalness for the Higgs mass.
A crucial issue however, which has been ignored in the entire discussion so far, is whether the very formulation
of the problem can be presented in a regularization-independent fashion.
As will now be seen, this actually turns out to be impossible.
In particular, within a dimensional regularization scheme, where all divergences appear as poles in the infinitesimal
parameter $\epsilon$, the problem is absent\footnote{Roughly, the role of $\epsilon^{-1}$ is similar to that of $\log \Lambda$
in a scheme with an explicit ultraviolet cutoff $\Lambda$. This thus means that there is no place for quadratic divergences
in a dimensional regularization scheme (see e.g. \citeN{tHooft1}).}.
To see this explicitly, note that the correction to the scalar mass, $\delta m^2$, to first order equals $-i \lambda G_F(0)_{\mbox{\scriptsize reg}}/2$,
where $G_F(0)_{\mbox{\scriptsize reg}}$ denotes a regularized Feynman propagator evaluated at zero momentum (cf. eq. (\ref{renormcond})).
In a dimensional regularization scheme, the coefficient of this first order term is thus given by
\begin{equation}
\frac{i \mu^{\epsilon} G_F(0)_{\epsilon}}{2} \; = \; \frac{\mu^{\epsilon}}{2(4 \pi)^{d/2}} \frac{\Gamma (1 - d/2)}{(m_0^2)^{1-d/2}} \; \: \stackrel{\mbox{\footnotesize $d = 4 - \epsilon$}}{=} \; \: \frac{m_0^2}{32 \pi^2}\left( - \frac{2}{\epsilon} \: + \: \gamma \: - \: 1 \: + \: \log \frac{m_0^2}{4 \pi \mu^2} \: + \: \mathcal{O}(\epsilon) \right)
\end{equation}
(with $\gamma \approx 0.57$ denoting the Euler-Mascheroni constant) and is thus \emph{proportional} to the (squared) bare
mass. This proportionality of the self-energy to $m_0^2$ is moreover preserved by higher-order contributions.
In other words, within a dimensional regularization scheme, radiative corrections cannot generate a nonzero mass if $m_0 =0$
and the custodial symmetry in this case is the scale invariance of massless $\lambda \phi^4$-theory\footnote{Cf. \citeN{Bardeen}.
Note that the mass parameter, $m$, in (\ref{phi4Lagr}) is the only dimensionful parameter of the full standard model 
Lagrangian, so that any scale invariance displayed within just the scalar field model is transposed to the full 
standard model. There are a few subtleties here, the most important one being that the scale invariance of the 
classical theory is anomalously broken (cf. footnote \ref{dimtrans}), but as it turns out, the scale anomaly is unrelated 
to the quadratic divergence of the scalar self-energy within an explicit cutoff regulator scheme and the latter represents
a separate, explicit breakdown of scale symmetry (cf. \citeANP{Bardeen}, ibid.).
It is not clear whether scale symmetry can still act as a custodial symmetry after including nonperturbative
effects or upon embedding the standard model into a more fundamental theory at short distances - although these
considerations evidently do not affect the standard model as such.}.\\
How, in view of this simple observation, can it be that the belief that there is a veritable Higgs naturalness problem
for the standard model is so widespread ?
Apart from the fact that the observation does not appear to be too well-known, the logical inference of the foregoing is 
that theorists who \emph{are} aware of it and still maintain that there is a problem of naturalness, have to resort to the
view that explicit cutoff schemes are in some way physically superior.
As far as I am aware however, this does not correspond to an actual position taken in the literature.
Regularization schemes that involve an explicit high-energy cutoff are certainly more physically intuitive 
than the dimensional scheme (at least if the cutoff is interpreted realistically and if the difficulties associated
with the EFT paradigm are ignored).
Yet the latter is in many ways mathematically superior, especially in the case of Yang-Mills theories.
Some further reflection along these lines then strongly suggests that a perspective in which priority is assigned to schemes
with an explicit ultraviolet cutoff is deeply problematic, since it would require an explicit demonstration that the entire 
empirical content of the standard model can be derived on the basis of such schemes.
Indeed, conventional wisdom has it that ``physics'' either should be or typically is independent of any particular 
chosen regulator, while in those (few) instants where it is not, this is generally due to a violation of some symmetry 
by one of the regulators, in which case the symmetry is to be taken as fundamental\footnote{\citeN{ItzZub} for instance 
state (p. 379) that ``the important point is that the final renormalized result does not depend on the choice of 
regularization [scheme]'', while \citeN{PesSch} point out (pp. 248-249) that the choice of regulator often has no effect 
on the predictions of the theory, but in those instants where it has, this is due to a violation of symmetry and ``in these
cases we take the symmetry to be fundamental and demand that it be preserved by the regulator''.}.
In the particular case at hand this would thus mean that dimensional regularization actually is to be preferred, 
as it does not explicitly break the classical scale invariance present within the massless theory.

\section{Discussion}\label{disc}

\noindent As a human achievement, the standard model of particle physics and its empirical verification - being the
outcome of a multitude of international collaborations involving tens of thousands of researchers - is an impressive 
construct, that is probably unrivalled in its kind.
As a scientific achievement however, the basis of the success of the standard model still awaits to be understood.
In particular, it remains to be seen how much of the model will be part of any future physics discourse - in the same sense, for 
instance, in which Newtonian physics, quantum theory and general relativity \emph{will} nearly certainly all remain part of a 
future physics discourse, despite the fact that these theories are not (or will very probably fail to be) fundamental theories.
Clearly, any ultimate explanation in terms of twenty-five elementary constituents - which corresponds to about half the 
number of chemical elements known to exist when Mendeleev proposed his periodic table, almost exactly a century prior to the birth of the 
standard model - seems highly implausible\footnote{It would moreover also misrepresent things considerably - since in
terms of elementary \emph{fields}, the standard model already contains fourty-five chiral fermion fields alone (i.e., corresponding to four 
left-handed doublets and seven right-handed singlets in each generation).}.
But does that mean that fundamental physics should once again embark on a journey to hunt down the putative truly elementary 
constituents of matter and interaction, in terms of which the entire physical universe can in principle be understood ?\\
During most of the four-decade old history of the standard model, it was generally held that the answer to this question 
was simply affirmative.
Furthermore, a fair amount of consensus even existed during this period that a leading (or even unique) candidate theory in
this regard consisted in the specific form of superstring/M-theory, with oscillating supersymmetric strings or higher-dimensional ``branes''
in ten (or eleven, twelve, $\cdots$) spacetime dimensions taking on the role of true atoms.
That is to say, leaving aside certain hypothetical duality features, these strings or branes were supposed to describe truly fundamental degrees of freedom (residing somewhere 
near the Planck scale) and all standard model particles should somehow be retrievable from particular string/brane states,
while it was moreover generally accepted that each spectrum of string/brane excitations should, as a rule, also contain a 
host of \emph{new} particles - in broad compliance with some particular ``low-energy'' projection of a supersymmetric GUT extension
of the standard model.
In fact, although no concrete contact with the physics of the standard model itself was ever made,
the apparent Higgs naturalness problem was generally interpreted as providing a major clue here, in the sense that if supersymmetry
was to resolve this (perceived) problem, superpartners of the standard model particles should definitely exist below or near 
the TeV scale\footnote{See e.g. \citeN{Polchinski} or \citeN{Kane}.}.
So far however, all searches for supersymmetric particles have come up empty-handed\footnote{\citeN{ATLAS2}, \citeN{CMS2}.}
and while this does not exclude the existence of such particles \emph{per se} (in fact, just as for the hallmark
proton decay of a typical GUT scenario, it appears that no foreseeable experiment could ever accomplish that),
there does appear to be a growing consensus that supersymmetry cannot be the answer to the (perceived) naturalness problem\footnote{See e.g. \citeN{Lykken}, who incidentally takes a similar position on the issue of naturalness.}.\\
But it was of course seen before that the belief that there exists a Higgs naturalness problem in the first place has
its roots in an incoherent use of renormalization theory. Consequently, there is no \emph{necessity}, on the basis of 
naturalness, for new physics to be residing near the TeV scale. This should give some pause for thought.
Even though, as noted, alternative solutions to the alleged naturalness problem have been pursued (most notably technicolour),
for many years, according to the dominant view in this regard, supersymmetry was simply the preferred option (in fact, at one point, 
the yet-to-be-built LHC was even portrayed as a ``superpartner factory'').
At a more fundamental level, things have not exactly played out for superstring theories and their later transmogrifications, 
as initially conveived of, either. Following the ``landscape crisis'' of about a decade ago, string apologists now largely 
seem to have abandoned the once high-held hope of a unique theory.
In fact, as a direct consequence of this crisis, string-inspired approaches have expanded their core business and now
postulate not merely entire new families of particles, but even some $10^{500}$ alternate universes. 
Fundamental physics has come a long way since the discovery of the positron back in 1928 !\\
These remarks of course do not prove the non-existence of a fundamental underlying theory positing a set of truly elementary
constituents.
However, they do entail that the obsession of high energy physics with such a theory during a large part of the twentieth century 
may have been fundamentally misplaced and that conceptually alternative approaches may well have a more realistic chance at success. 
It seems that an important reason why such an option is nevertheless not welcomed with great enthusiasm at present is 
that the notion of \emph{atomist} reductionism is often conflated with that of reductionism more generally.
But as already pointed out by Weinberg nearly a quarter of a century ago, the present focus on elementary particle physics - essentially 
atomist reductionism - as the seemingly best way to uncover the final laws of physics is an incidental aspect of 
reductionism which may change in the future\footnote{Cf. \citeN{Weinberg7}.}.
In compliance with general scientific methodology, what is crucially needed in order to advance fundamental physics is the uncovering
of the general \emph{principles} underlying Nature's workings at the deepest possible level (in this regard, philosophies 
based on a principle of naturalness or superstring least action, even though inapt in advancing fundamental physics so far, 
certainly are \emph{prima facie} legitimate approaches to meet that objective).\\
Evidently, many possible approaches towards a principles theory are a priori imaginable and in the absence of a
definite agreed theory, any of these should be encouraged to develop further.
One particular broad approach, as recently advocated by the author, is based on assigning 
a key guiding role to (certain) observational principles, in close recognition of the actual history of relativity theories\footnote{Cf. \citeN{Holman3}. Even though the case presented here strictly
speaking only pertains to the ``quantum gravity'' problem, the discussion can evidently be adapted to the present context without
difficulty, should that indeed be necessary.}.
In the specific case of quantum gravity, this approach furthermore vaguely hinted at a particular mathematical 
formalism, namely twistor theory, and it is interesting to observe that this feature of it could
well render it complementary to another recently promoted approach, based on a postulated local
conformal invariance of physical laws\footnote{See e.g. \citeN{Mannheim}, \citeN{tHooft5}, \shortciteN{AmArGuMa}.
The connection with twistor theory arises because twistors can be characterized as reduced spinors for $\mbox{O}(4,2)$ 
(i.e., the group of linear isometries in a six-dimensional flat linear space with a metric of signature $(+,+,+,+,-,-)$)
and the latter group is $2-1$ isomorphic to the conformal group, $C(3,1)$ of Minkowski spacetime.
Alternatively, if a dimensional reduction to 2 is contemplated in the far-ultraviolet region, the causal structure 
reduces to that of an ordinary flat spacetime (so that gravity is indeed ``switched off'') and the local Lorentz 
geometry can be Wick rotated to a Euclidean one, which can subsequently be identified with a Riemann sphere 
using standard stereographic projection.
As it happens however, the celestial ``sphere of vision'' of an observer at a point in ordinary Minkowski spacetime
can also be viewed as a Riemann sphere, with conformal transformations on the latter precisely representing
(restricted) Lorentz transformations in Minkowski spacetime.
In twistor theory, the Riemann spheres arising in this manner are basically taken to (nonlocally) \emph{represent} ordinary
spacetime points (the analogy with the dimensionally reduced framework is thus somewhat formal, as there are
still ``four dimensions worth'' of Riemann spheres in the twistor description of things, rather than two).}.
Although it seems at first glance difficult to see how the notion of conformal invariance could play any
role in a fundamental description of things (in sharp contrast to the approximate Poincar\'e invariance of spacetime, 
phenomenology is evidently grossly non-conformally invariant), there are in fact several reasons to suspect that 
such a notion could well be crucial for future theory development.\\
First, the logic of the standard model, if taken seriously, simply entails that at very high energies, physical laws
become conformally invariant. That is to say, even though the standard model contains precisely one dimensionful
parameter (essentially the parameter $m$ in (\ref{phi4Lagr}); cf. the accompanying note), this parameter does not 
acquire a clear physical interpretation until \emph{after} electroweak symmetry breaking (when it essentially
becomes proportional to the inverse square root of Fermi's constant, $G_{\mbox{\scriptsize F}}$).
In particular, since all massive particles acquire their masses in the transition to the Higgs phase according to the 
standard model, ``before'' this transition (which is usually portrayed as an actual physical transition occurring in 
the very early universe - approximately at $\tau_{\mbox{\tiny EW}} \simeq 10^{-12} \, \mbox{s.}$), only massless particles 
existed and spacetime geometry was essentially conformal.\\
A second argument for conformal invariance enters at a much more elementary level and in fact goes back to the early history 
of general relativity.
As is well known, only three years after Einstein completed the formulation of his new theory of gravitation and spacetime,
an interesting alternative theory, based on a postulated invariance with respect to local rescalings of the spacetime
metric, was proposed by \citeN{Weyl1}. Today, Weyl's theory is generally regarded as incompatible with observations
and, consequently, of interest only for historical purposes (not in the least also because the whole notion of \emph{gauge symmetry},
according to its modern usage, is widely taken to have originated in the \citeyearNP{Weyl2} sequel of the theory). 
It appears to be somewhat less well recognized however that the inclusion of local conformal invariance in the original 1918 
theory was a very natural step, which merely took an important part of the guiding philosophy behind general relativity
to its logical conclusion.
Indeed, in a truly local (i.e., ``infinitesimal'') geometry, it should, without any further information, not only make no sense to compare the
\emph{directions} of two vectors at points separated by a finite distance, but this should also be true for the
\emph{magnitudes} of those vectors.
The trouble with this idea, as famously pointed out by Einstein soon afterwards, was that the general path-dependence 
of proper time or length thereby implied clearly contradicted empirical evidence - such as the existence of stable atomic spectra or 
rods of fixed lengths.
However, it again appears to be somewhat less well appreciated that Weyl's response to this objection - namely that the 
actual behaviour of physical clocks or measuring rods would have to follow from a dynamical theory of matter - although
seemingly rather contrived, strictly kept open the possibility of later vindication of the theory.
In particular, given that a consistent theory of quantum gravity is likely to involve some subtle, but nevertheless 
profound modifications to both quantum theory and general relativity, it is very well conceivable that, when such a theory is finally
obtained, it will somehow manage to avoid the above-noted anomaly in practice.
At present, this of course remains pure speculation.
Nevertheless, a highly attractive (especially from a particle physics perspective), well-known feature of Weyl's 
conformally invariant theory so far unmentioned, is that it near-automatically incorporates a description of electromagnetism.
In other words (and as apparently not anticipated by Weyl), the requirement of exact local conformal invariance
almost unavoidably leads to a \emph{unified theory} of gravitation and electromagnetism\footnote{It should be remarked here
that Weyl made various proposals for a specific form of the action over the years, but did not seem to have settled on
a final choice. See \citeN{Holman2} and references therein for further discussion.}.\\
Finally and related to the previous motivation, within the context of general relativity proper, it is the \emph{conformal}
part of the Riemann curvature, i.e., the Weyl tensor, $C_{abc}^{\mspace{24mu}d}$, which most appropriately represents the gravitational
degrees of freedom within the theory (the Ricci tensor, $R_{ab}$, together with the scalar curvature, $R$, are effectively equivalent
to the stress-energy tensor via Einstein's equation and thus correspond more appropriately to non-gravitational ``source degrees of
freedom'').
One could thus say that, as far as ``gravity just by itself'' is concerned, general relativity actually displays a basic
local conformal invariance, while the coupling of gravity to non-gravitational matter and fields within the theory somehow
breaks this invariance.\\
Whether the notion of conformal invariance will indeed fulfill the role here envisaged of course remains to be seen. 
The key point of the foregoing disgression however is that unorthodox approaches to a principles theory for physics 
beyond the standard model can and indeed should be pursued.
The particle physics standard model has been characterized as a ``Shakespearean king disguised as a beggar''.
Now that it appears that all its elementary constituents may, for the first time in the model's history,
have been identified experimentally, to find out whether the king has diamonds in his pockets would seem timely.\footnote{Cf. \citeN{Connes}.}.


\bibliographystyle{eigen}
\bibliography{higgsnat}

\end{document}